%% file: main.tex
\documentclass[lettersize,journal]{IEEEtran}
\usepackage{amsmath, amsfonts, amssymb, amsxtra, mathrsfs, pifont}
\usepackage{algorithmic}
\usepackage{array}
\usepackage[caption=false]{subfig}
\usepackage{textcomp}
\usepackage{stfloats}
\usepackage{url}
\usepackage{verbatim}
\usepackage{graphicx}
\usepackage{xcolor}
\usepackage[nolist]{acronym}
\usepackage{accents}
 
\def\BibTeX{{\rm B\kern-.05em{\sc i\kern-.025em b}\kern-.08em
    T\kern-.1667em\lower.7ex\hbox{E}\kern-.125emX}}
\usepackage{balance}

\makeatletter
\newcommand{\ostar}{\mathbin{\mathpalette\make@circled\star}}
\newcommand{\make@circled}[2]{%
\ooalign{$\m@th#1\smallbigcirc{#1}$\cr\hidewidth$\m@th#1#2$\hidewidth\cr}%
}
\newcommand{\smallbigcirc}[1]{%
  \vcenter{\hbox{\scalebox{0.77778}{$\m@th#1\bigcirc$}}}%
}
\makeatother  
\newcommand{\re}[1]{{\mathfrak{Re}}\left\{#1\right\}}

\newcommand{\diag}[1]{{\mathrm{diag}}\left\{#1\right\}}
\newcommand{\hathat}[1]{\hat{\vphantom{\rule{1pt}{5.5pt}}\smash{\hat{#1}}}}

\newcommand{\secref}[1]{Section \ref{#1}}
\newcommand{\tabref}[1]{Table \ref{#1}}
\newcommand{\figref}[1]{Fig. \ref{#1}}

\begin{acronym}[ACRONYM]
\acro{3GPP}{the third generation partnership project}
\acro{5G}{fifth generation}
\acro{6G}{sixth generation}
\acro{ADC}{analog-to-digital converter}
\acro{AI}{artificial intelligence}
\acro{AoA}{angle-of-arrival}
\acro{AoD}{angle-of-departure}
\acro{BD}{backscatter device}
\acro{BER}{bit error rate}
\acro{BPSK}{binary phase shift keying}
\acro{BS}{base station}
\acro{CFO}{carrier frequency offset}
\acro{CIR}{channel impulse response}
\acro{CFR}{channel frequency response}
\acro{D-MIMO}{distributed multiple-input multiple-output}
\acro{EM}{electromagnetic}
\acro{FFT}{fast Fourier transform}
\acro{GDOP}{geometric dilution of precision}
\acro{GNSS}{global navigation satellite system}
\acro{ICI}{inter-carrier interference}
\acro{IFFT}{Inverse Fast Fourier Transform}
\acro{IQ}{in-phase and quadrature}
\acro{JCS}{joint communication and sensing}
\acro{JRC}{joint radar and communication}
\acro{ICI}{inter-carrier interference}
\acro{IoT}{Internet of Things}
\acro{LFM}{linear frequency modulation}
\acro{LoS}{line-of-sight}
\acro{MIMO}{multiple-input multiple-output}
\acro{ML}{maximum likelihood}
\acro{mmWave}{millimeter-wave}
\acro{MSps}{megasamples per second}
\acro{MUSIC}{MUltiple SIgnal Classification}
\acro{NLoS}{non-line-of-sight}
\acro{NMAE}{normalized mean absolute error}
\acro{NR}{new radio}
\acro{OFDM}{orthogonal frequency-division multiplexing}
\acro{PLL}{phase locked loop}
\acro{PRS}{positioning reference signal}
\acro{QAM}{quadrature amplitude modulation}
\acro{QoS}{Quality of Service}
\acro{RAN}{radio access network}
\acro{RAT}{radio access technology}
\acro{RF}{radio frequency}
\acro{RFID}{radio frequency identification}
\acro{RedCap}{reduced capacity}
\acro{RTK}{real-time kinematic}
\acro{RTT}{round-trip-time}
\acro{SLAM}{simultaneous localization and mapping}
\acro{SNR}{signal-to-noise ratio}
\acro{SINR}{signal-plus-interference-to-noise ratio}
\acro{ToA}{time-of-arrival}
\acro{TDoA}{time-difference-of-arrival}
\acro{TR}{time-reversal}
\acro{TRP}{transmission and reception point}
\acro{TXRX}[TX/RX]{transmitter/receiver}
\acro{TX}{transmitter}
\acro{RX}{receiver}
\acro{RMSE}{root mean squared error}
\acro{UE}{user equipment}
\acro{multi-RTT}{multi-cell round-trip-time}
\acro{UL-TDOA}{uplink time-difference-of-arrival}
\acro{DL-TDOA}{downlink time-difference-of-arrival}
\acro{UHF}{ultrahigh frequency}
\acro{UMi}{3D-urban micro}
\acro{UMa}{3D-urban macro}
\acro{UWB}{ultrawideband}
\acro{FR1}{frequency range 1}
\acro{FR2}{frequency range 2}
\acro{CDF}{cumulative distribution function}
\acro{ToF}{time-of-flight}
\acro{CSI}{channel state information}
\acro{OT}{operational technology}

\end{acronym}
\def\T{\top}
\def\H{\mathrm{H}}

\begin{document}

\title{Next-Generation Backscatter Networks for Integrated Communications and RF Sensing}
\author{Traian E. Abrudan, Kartik Patel, John Kimionis, Tara Esmaeilbeig, Eleftherios Kampianakis, Sahan Damith Liyanaarachchi, and Michael Eggleston 
\thanks{T.~E. Abrudan is with Nokia Bell Labs, Espoo, Finland. K. Patel, J. Kimionis, T. Esmaeilbeig, E. Kampianakis, and M. Eggleston are with Nokia Bell Labs, Murray Hill, NJ, USA. S.~D. Liyanaarachchi is with Nokia, Tampere, Finland. Email addresses: \texttt{\{traian.abrudan, kartik.patel, ioannis.kimionis, lefteris.kampianakis\}@nokia-bell-labs.com, \{tara.esmaeilbeig, sahan.liyanaarachchi\}@nokia.com}}}



\maketitle
\begin{abstract}
This paper provides a comprehensive analysis and theoretical foundation for next-generation backscatter networks that move beyond communication and integrate RF location sensing capabilities. An end-to-end system model for wideband OFDM backscatter systems is derived, including detailed characterization of propagation channels, receiver chain impairments, RF tag operation, and unsynchronized network nodes. The theoretical system model is validated through experimental evaluation using actual hardware, demonstrating the detailed model's accuracy. A practical bistatic ranging method that can operate with unsynchronized nodes is presented, along with the Cramér-Rao Lower Bound (CRLB) derived to show the achievable performance limits. Our experimental results demonstrate the system performance for communication, RF sensing, and ranging, while also benchmarking against the derived theoretical limits. This analytical framework and experimental validation establish fundamental understanding of distributed, unsynchronized backscatter systems for future machine-type communication networks that are deployed in massive scale, while remaining energy-efficient.
\end{abstract}
\section{Introduction}
The vision of connecting trillions of \ac{IoT} devices dictates a paradigm shift in how we approach device complexity, power consumption, and network scalability. This is especially relevant as we move towards next-generation networks, from the 5G era into the 6G era, where energy efficiency will be crucial for massive-scale, wide-area system deployments. Conventional active radio, while powerful, often proves prohibitively energy-intensive and costly for ultra-dense deployments of sensors, wireless IDs, or tracking devices. Therefore, these types of systems are in need of different types of machine-type communication in order to be integrated in 6G-and-beyond networks.
Backscatter communication, long established in commercial \ac{RFID} systems \cite{2005_Land,2007_ChaHa}, has been proposed as a promising solution to address these challenges, by conveying sensor information via signal reflection rather than active radiation \cite{2008_VanBleLei,2014_KamKimTou}. 
Reflection communication requires simplified RF front-ends that modulate and reflect over-the-air signals. By eliminating the need for complex RF components and local oscillators for carrier generation, backscatter technology enables ultra-low-power operation and simplified device architecture, compared to active radio counterparts. 

While historically confined to short-range identification applications, advances in bistatic backscatter topologies \cite{2014_KimBleSah,2016_VouDasBle} and high-gain tag architectures \cite{2018_AmaTorDur,2014_KimGeoTen} have extended the operating ranges of such systems beyond their traditional limits. Additionally, there has been high interest within the research community to study these systems as dynamic information communication systems and characterize their performance, capacity limits, and network capabilities \cite{2017_HanHua,2017_LuJiaNiy,2016_WanGaFan}. Backscatter device readouts can be implemented either with custom protocols \cite{2021_KarSafSmi}, by re-using existing wireless infrastructure (e.g., BLE and WiFi \cite{2016_ZhaRosHuGan,2016_ZhaBhaJos,2023_MenDunKuoYu}), or even with newly forming standards (e.g., 3GPP Ambient IoT \cite{2024_ButManPra}). Distributed networks can be formed for ultra-high density multiple access of backscatter devices \cite{2025_PatZhaKim} attached to objects, people, machines, or vehicles in different environments, offering scalable and energy-efficient solutions for Industry 4.0 and 5.0 \ac{OT}.


This paper expands on the potential of backscatter technology from simple identification tags that convey low-rate information to sophisticated communication and sensing platforms that offer wide-area intelligence of \textit{what} and \textit{where} an object is. It offers the theoretical foundation for end-to-end backscatter systems, which not only enable their use in next-generation networks with highly-flexible \ac{OFDM} communication signaling but also enable a powerful RF sensing mechanism. 
It presents a detailed system model that includes the propagation channels, receiver chains, signal impairments, and RF tag operation. The model also captures the time, phase and frequency offsets due to non-synchronized system nodes and characterizes the performance of the backscatter system from both a communication and sensing point of view. The model is backed by experimental validation of the system with actual hardware, proving the validity of the model, and justifying the extensive description of the large parameter space that these systems encompass. In addition, this paper analyzes a wireless sensing scenario where the system performs backscatter device \textit{ranging}. To define the theoretical limits of ranging performance, it first presents the Cramér-Rao Lower Bound (CRLB). 
It then introduces a \ac{ToF}-based ranging method for backscatter device ranging, a highly desired method for positioning systems but typically used with active, power-demanding devices with tight time synchronization, e.g., \ac{UWB}\cite{2021_LaaUlpMuh}. 
This paper shows the feasibility of the \ac{ToF}-based ranging on backscatter devices with experimental testbed, while maintaining the fundamental goals of low-complexity and ultra-low power consumption that make backscatter systems attractive for large-scale deployments.


\subsection{Related work}

A large-area RF sensing system requires long-range backscatter tag reception, far beyond conventional RFID system operating ranges, that are limited to a few meters. Work in \cite{2014_KimBleSah,2016_VouDasBle} has demonstrated 100+ meter backscatter signal reception, by utilizing a distributed transmitter-receiver topology, known as \textit{bistatic} backscatter. In these systems, the separation of TX-RX pairs offers two degrees of freedom for area coverage (TX-tag range and tag-RX range), as opposed to conventional \textit{monostatic} RFID systems which are limited by the round-trip signal path loss (Reader-tag range). Our work utilizes the bistatic architecture which significantly increases the flexibility of infrastructure deployment and communication/sensing area coverage for backscatter tags.

RF Sensing and multicarrier ranging utilize large bandwidth values to increase resolution. Large illumination bandwidth can be achieved with an \ac{OFDM} carrier/multicarrier signal, on top of which the backscatter tag will modulate information. Backscattering on top of OFDM illumination signal was shown in \cite{2023_AleVouBle}, where \ac{ML} joint and disjoint detectors for illuminator symbol and tag symbol detection were derived. OFDM illumination was also shown in \cite{2023_LiaWanRutJan}, where LTE reference signals were used. In our work, we also employ OFDM, but we couple it with a large frequency shift to achieve  high dynamic range reception at the receiver.

Frequency-shifting was employed in \cite{2016_ZhaRosHuGan,2018_VouBle,2020_DinLihRom} to avoid the interference of the backscattered signal spectrum with the illumination signal spectrum, to aid communication. In our work, we expand the usefulness of the large frequency shift to achieve spatial-frequency diversity, i.e., effectively observe a wireless channel through a backscatter tag at different bands, effectively transforming it into a frequency division duplexing system. Being able to sense the amplitude and phase differences between two or more bands (lower and/or upper sidebands) in turn enables wideband \textit{ranging} of the backscatter tag. 


Work in \cite{2019_QiAmaAlhDur, 2021_QiAmaAlhDur} demonstrates experimentally a monostatic backscatter phase-based ranging system, where frequency-hopping of continuous wave (CW) carriers is utilized. Our system prioritizes instantaneous channel state information extraction, rather than multi-point data collection over time and therefore relies on wideband illumination. OFDM, used in 5G, 5G-Advanced, and most likely 6G, offers high flexibility in setting up the number of subcarriers, spacing, effective bandwidth, etc. to generate the multi-carrier TX signal, which in turn translates to a multi-carrier OFDM backscatter signal observed through  separate channels.
%
%
%
%

%
%
%
%
%
%
%
%
\subsection{Contributions}
This work's main contributions can be summarized as follows.
\begin{itemize}{}{}
    \item Full OFDM Backscatter system design with intrinsic RF sensing and ranging capabilities, in addition to communication capabilities. 
    \item Comprehensive end-to-end system modeling, including detailed description of transmitted waveforms, RF tag operation, channel effects, receiver impairments, fully taking into account the non-synchronized nature of the system. System analysis is offered in both the time- and frequency domains and can be the foundation for future research on such systems.
    \item Detailed derivation of analytic expressions for the receiver-estimated radio channels in a distributed/bistatic architecture, which is crucial for RF sensing operations and accurate environment perception.
    \item Description of ranging methods in a non-synchronized system and in the presence of multipath, completed with the derivation of the Cramér-Rao Lower Bound for bistatic ranging with frequency-shifted backscatter.
    \item Experimental validation of the large parameter space model and characterization of communication performance with \ac{BER} metric, RF sensing performance with \ac{RMSE} quantification between expected and estimated \ac{CSI}, as well as ranging performance with ranging accuracy \ac{CDF}.
\end{itemize}
\subsection{Notation}
Throughout this paper, bold-weight font is used to represent matrices and vectors, and regular-weight font to represent scalars. Frequency-domain quantities are further represented by calligraphic font. $(\cdot)^\T$ and $(\cdot)^\text{H}$ denote the transpose and the Hermitian transpose of a matrix, respectively. Identity matrix, zero vector and ones vector of dimension $p$ are denoted by ${\mathbf I}_p$, ${\mathbf 0}_p$ and $\mathbf{1}_p$, respectively. Rows and columns of a matrix are indexed from $i_1$ to $i_2$ using $[i_1:i_2]$ notation. Element-wise multiplication and division of vectors are denoted by $\odot$ and $\oslash$, respectively, whereas linear and circular convolution operations are denoted by $\star$ and $\ostar$, respectively. Modulo operation is denoted by $\oplus$.  Real part of a complex-valued quantity is denoted by $\re{\cdot}$, and $\jmath = \sqrt{-1}$ denotes the imaginary unit. Rectangular window of duration $T$ is denoted by $\text{rect}_{[0,T)}(t)$. The statistical expectation w.r.t. a random variable $X$ is denoted by ${\mathbb E}_X\{\cdot\}$.

\section{Backscatter System Model}
\begin{figure}[t]  
\centering
\includegraphics[width=0.8\columnwidth]{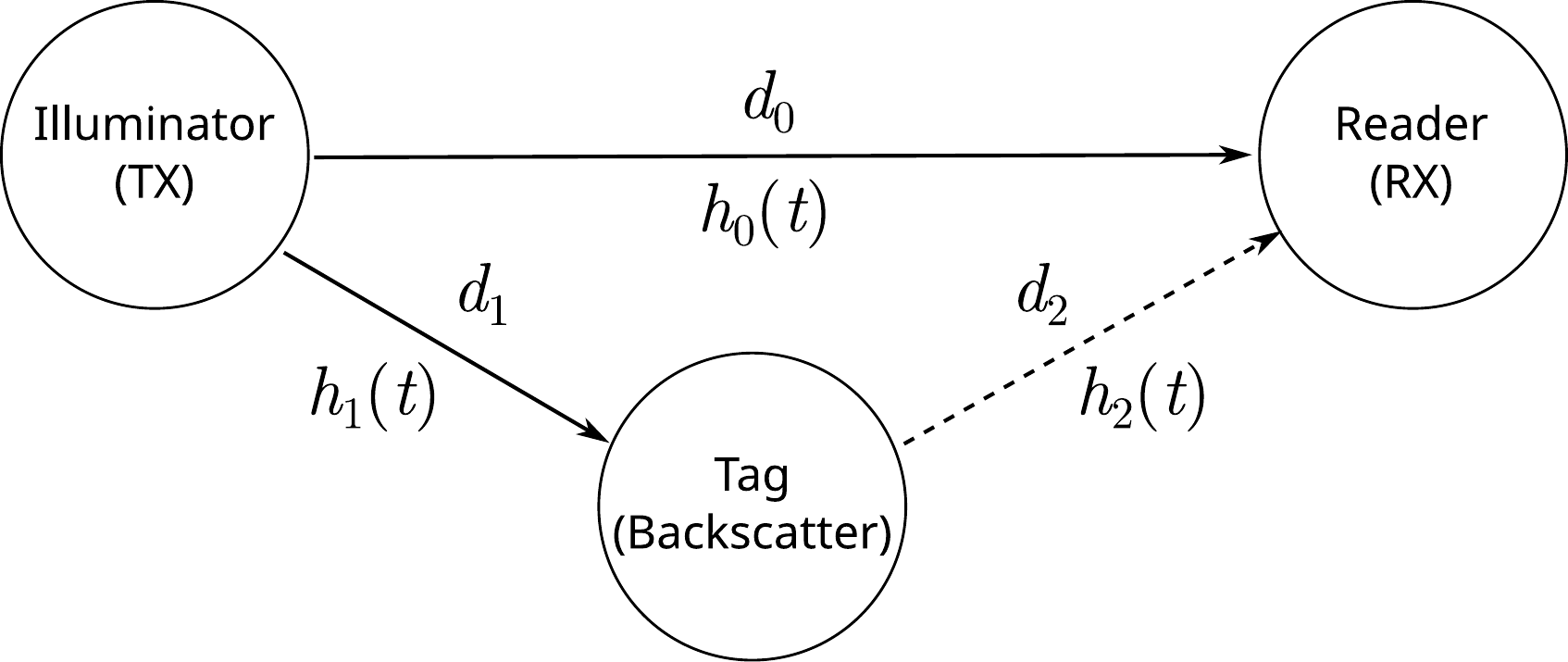} 
\caption{The bistatic backscatter system model. Solid lines denote active transmission channels (TX--RX and TX--Tag), the dashed line is the backscatter (reflection) channel (Tag--RX). \label{fig:system_block_diagram} }
\end{figure}

In this paper, we provide the full system model of a wideband OFDM bistatic backscatter system which is illustrated in \figref{fig:system_block_diagram}. The system nodes are similar to the ones in \cite{2014_KimBleSah}, while the signaling is based on wideband OFDM, instead of single-carrier narrowband illumination and backscattering. The illuminator, or \ac{TX} broadcasts its own multicarrier signal. The \ac{BD}, referred from now on as the tag, not only reflects and modulates the illumination signal, but also frequency-shifts it away from the illumination carrier, and into an adjacent channel, similarly to \cite{2016_ZhaRosHuGan}.
Since the illumination and the backscatter signals are in different frequency channels, the receiver avoids being blinded by the illuminator through channelizing (with frequency selective filtering). This increases the operating dynamic range, and therefore achieves longer reception range of the backscatter signal. Then, the reader, or \ac{RX} processes the corresponding signals in a per-band basis. 
The high-level system diagram in \figref{fig:system_block_diagram} shows the main blocks: Illuminator (TX), Tag (backscatter device), Reader (RX). The channels between the nodes are denoted as $h_i(t)$ and the respective separation distances between the nodes are denoted as $d_i$. Solid lines denote active transmission channels (TX--RX and TX--Tag), whereas the dashed line is the backscatter (reflection) channel (Tag--RX).
\subsection{OFDM Illumination}
\input{illuminator}
\subsection{Bistatic channel model}
\input{channel}

\subsection{Tag Operation}
\input{tag}
\subsection{Multi-band Reader}\label{sec:reader}
\input{reader}

\section{Bistatic ranging}\label{sec:ranging}
\input{ranging}

\section{Experiments}\label{sec:results}
\input{results}

\section{Conclusion}
This paper presented a complete theoretical system model backed by experimental validation for wideband bistatic backscatter networks that can integrate communication and RF sensing capabilities. A detailed end-to-end system model for wideband OFDM backscatter was derived, accounting for propagation channels, receiver chain impairments, RF tag operation, and unsynchronized network nodes. The analytical framework was validated through careful experimental evaluation using actual hardware, demonstrating excellent agreement between theoretical and experimental CSI estimation, with \ac{NMAE} below 0.035 for TX-RX (active) channels and 0.12 for TX-Tag-RX (backscatter) channels. The communication performance was assessed and the resulting \ac{BER} vs SNR relation matches the theoretical expected one. The bistatic backscatter ranging performance was characterized experimentally and was benchmarked against the derived CRLB.

The comprehensive analytical expressions and experimental validation presented in this work establish the foundation for future research in distributed, unsynchronized backscatter systems. This framework enables the development of sophisticated ranging methods for multipath environments and paves the way for large-scale deployment of energy-efficient machine-type communication networks that seamlessly integrate sensing capabilities.
Future research can focus on developing advanced algorithms to overcome the current limitations in multipath environments and further close the gap between practical implementation and theoretical performance bounds.

\bibliographystyle{IEEEtran}
\bibliography{refs}

\newpage
\appendices
\section*{Cramér–Rao Lower Bound Derivation}\label{app:crlb}
\input{crlb}
\end{document}

%% file: illuminator.tex
We assume an \ac{OFDM} illumination signal with $N$ subcarriers 
equally spaced at $\Delta F$ that span a bandwidth $B = N\Delta F$. Without loss of generality, we assume the set of active subcarriers is 
\begin{equation}
{\mathbb A} = \left\{-\frac{N-1}{2},\ldots,+\frac{N-1}{2} \right\}.
\end{equation}
The complex-valued frequency-domain \ac{OFDM} illumination symbols, stacked in an $N \times 1$ column vector, are denoted by 
\begin{equation}
\boldsymbol{\mathcal S} = [{\mathcal S}_{-(N-1)/2}, \ldots, {\mathcal S}_{(N-1)/2}]^\T. \label{eq:ofdm_fd_vector}
\end{equation}

Let ${\mathbf F}_N$ be the $N \times N$ unitary \ac{FFT} matrix in which the $(k,n)$-th entry is given by
\begin{equation}
\{{\mathbf F}_N\}_{k,n}=\frac{1}{\sqrt{N}}\exp\Big(-\jmath \frac{2\pi k}{N}n\Big). \label{eq:FFT_matrix}
\end{equation}
The corresponding discrete time-domain output vector is
\begin{equation}
{\mathbf s} = [s_0, \ldots, s_{N-1}]^\T = {\mathbf F}_N^\textrm{H} \boldsymbol{\mathcal S}, \label{eq:ofdm_discrete_sig}
\end{equation}
where $s_k,~k=0,\ldots,N-1$ are the time-domain \ac{OFDM} samples that are repeated at the illuminator\footnote{Due to the \ac{OFDM} symbols repetition, cyclic prefix may be omitted. If cyclic prefix is used, the rest of the analysis still holds, because of the circular convolution properties on repeating OFDM symbols with (same) cyclic prefix.} in a serialized manner. The samples are then converted into an analog signal $s(t)$ that passes through the TX analog front-end characterized by the impulse response $g^\text{TX}(t)$. The transmitted baseband signal is
\begin{equation} 
s^\text{TX}(t) = s(t) \star g^\text{TX}(t). \label{eq:tx_analog_bb}
\end{equation}
The transmitted baseband signal is then up-converted by mixing with the carrier wave 
\begin{equation}
c^\text{TX}(t) = \exp \left[ +\jmath (2\pi F_\text{c}t + \phi_{\text{TX}}) \right],\label{eq:tx_carrier}
\end{equation}
with carrier center frequency $F_\text{c}$ and initial phase $\phi_{\text{TX}}$. 
The resulting up-converted waveform is 
\begin{equation} \label{eq:upconv_ofdm}
u(t) = \re{s^\text{TX}(t) c^\text{TX}(t)}.
\end{equation}
The signal $u(t)$, called \textit{illumination signal}, is transmitted by the illuminator.

%% file: channel.tex
\begin{figure}
\centering
\includegraphics[width=.8\columnwidth]{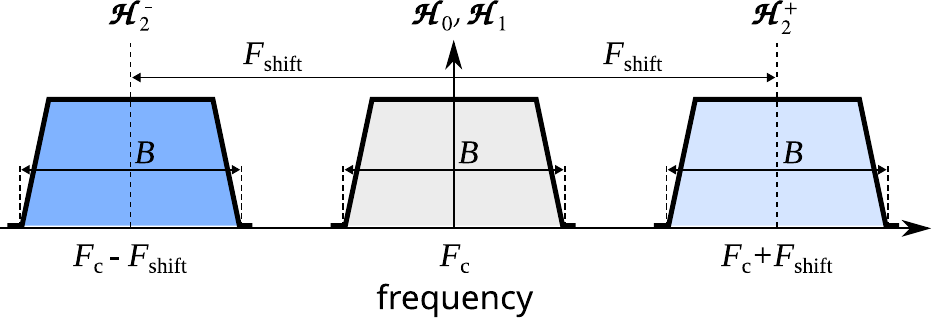} 
\caption{Overview of the different channels in frequency domain. 
Channel $\boldsymbol{\mathcal H}_0$ corresponds to the TX-RX link, and is centered at the illuminator carrier frequency $F_\text{c}$. 
Channel $\boldsymbol{\mathcal H}_1$ corresponds to the TX-tag link, and is also centered at the illuminator carrier frequency $F_\text{c}$.
Channels $\boldsymbol{\mathcal H}_2^-$ and $\boldsymbol{\mathcal H}_2^+$ correspond to the tag-RX link, and they are shifted from the illuminator carrier frequency at $F_\text{c} - F_\text{shift}$ and $F_\text{c} + F_\text{shift}$, respectively. 
\label{fig:channel_spectrum} }
\end{figure}

In this section, we provide the {\em baseband-equivalent channel model} for each link (See \figref{fig:channel_spectrum}):
\begin{itemize}
\item {\bf Channel $0$}, the direct TX-RX channel: the channel centered at the illuminator carrier frequency $F_\text{c}$ whose \ac{CIR} is $h_0(t)$, and whose \ac{CFR} at the illuminator subcarrier frequencies is described by the vector $\boldsymbol{\mathcal H}_0$.
\item {\bf Channel $1$}, the TX-tag channel: the first segment of the compound channel of interest (TX-tag-RX), also centered at the illuminator carrier frequency $F_\text{c}$ whose \ac{CIR} is $h_1(t)$, and whose \ac{CFR} $\boldsymbol{\mathcal H}_1$.
\item {\bf Channel $2$}, the tag-RX backscatter channel: the second segment of the compound channel of interest (TX-tag-RX) whose \ac{CIR} is $h_2(t)$. In frequency domain, the backscatter signal occupies two different sub-bands shifted down/up from the illuminator carrier frequency by $\mp F_\text{shift}$, respectively: 
    \begin{itemize}
    \item {\bf Channel $2-$}, the lower band channel centered at $F_\text{c} - F_\text{shift}$ whose \ac{CFR} is $\boldsymbol{\mathcal H}_2^-$,
    \item {\bf Channel $2+$}, the upper band channel centered at $F_\text{c} + F_\text{shift}$ whose \ac{CFR} is $\boldsymbol{\mathcal H}_2^+$.
    \end{itemize}
\end{itemize}

Next, we provide a common description of Channels 0, 1 and 2, pointing out the differences.
We assume the TX-RX distance $d_0$, the TX-tag distance $d_1$, and the tag-RX distance $d_2$, respectively, as in \figref{fig:system_block_diagram}. Therefore, the propagation times along the corresponding \ac{LoS} paths are $t_i=d_i/c, \: i \in \{0,1,2\}$, where $c$ is the speed of light. In addition to the \ac{LoS} path, we also assume that there are multiple additional \ac{NLoS} paths whose path lengths are denoted by $d_i^l$, and whose propagation delays are $t_i^l = d_i^l/c$. 
We further assume that each path has the associated complex-valued path-loss coefficient $\alpha_i^l$.\footnote{For consistency of notation, we assume that for the ``first path", $d_i^0=d_i$, $t_i^0=t_i$ and $\alpha_i^0=\alpha_i$.}
Then, the continuous-time \ac{CIR} can be represented by a function $h_i(t)$ comprised of $L_i$ discrete paths. 
\begin{equation} 
h_i(t) = \sum_{l=0}^{L_i-1} \alpha_i^l \delta(t-t_i^l), \qquad t_i^0 \leq t_i^1 \leq \enspace \cdots \enspace \leq t_i^{L_i-1}, \label{eq:multipath_ch0_imp_resp}
\end{equation}
where superscript $l=0$ corresponds to the \ac{LoS} path.
The corresponding \ac{CFR} of the $i$th channel ($i \in \{0,1,2\}$) is
\begin{equation}
\mathcal{H}_i(F) = \sum_{l=0}^{L_i-1} h_i(t_i^l) \exp(-\jmath 2 \pi F t_i^l), \quad i \in \{0,1,2\}.  \label{eq:multipath_ch0_freq_resp}
\end{equation}
Let us define the discrete-time propagation delay (in sampling periods) for the $l$-th propagation path seen at baseband with sampling period $T_\text{samp}$
\begin{equation} \label{eq:multipath_ch0_discrete_delay}
\tau_i^l = \frac{t_i^l}{T_\text{samp}}, \quad l = 0, \ldots, L_i-1, \quad i \in \{0,1,2\}.
\end{equation}
We are only interested in the \ac{CFR}s $\mathcal{H}_i(F)$ at the $n$th subcarrier in each band ($n \in {\mathbb A}$), as follows:
\begin{align} 
\mbox{Channels}~ 0, 1: \quad F_n &= F_\text{c}+n\Delta F,  \label{eq:los_ch01_discrete_F} \\
\mbox{Channels}~2\mp: \quad F_n^\mp &= F_\text{c} \mp F_\text{shift} + n\Delta F. \label{eq:los_ch2-+_discrete_F} 
\end{align}
Combining \eqref{eq:multipath_ch0_freq_resp}, \eqref{eq:multipath_ch0_discrete_delay} with \eqref{eq:los_ch01_discrete_F}, \eqref{eq:los_ch2-+_discrete_F}, respectively, and considering that $\Delta F=B/N$ and the sampling frequency $F_\text{samp}=1/T_\text{samp}$, we obtain the \ac{CFR}s at the illuminating ODFM subcarriers of each band:
\begin{align} 
{\mathcal H}_i(n) &= \sum_{l=0}^{L_i-1} \alpha_i^l \exp\left[-\jmath 2\pi \left(\frac{F_\text{c}}{F_\text{samp}} + \frac{n}{N} \right) \tau_i^l \right], \quad i \in \{0,1\}, 
\label{eq:multipath_ch01_discrete_freq_resp_tau_l} \\
{\mathcal H}_2^\mp(n) &= \sum_{l=0}^{L_2-1} \alpha_2^l \exp\left[-\jmath 2\pi \left(\frac{F_\text{c}\mp F_\text{shift}}{F_\text{samp}} + \frac{n}{N} \right) \tau_2^l \right].\label{eq:multipath_ch2_discrete_freq_resp_tau_l} 
\end{align}
Collecting the \ac{CFR}s at the subcarrier frequencies $n \in {\mathbb A}$ in each band in an $N \times 1$ vector, \eqref{eq:multipath_ch01_discrete_freq_resp_tau_l}--\eqref{eq:multipath_ch2_discrete_freq_resp_tau_l} can be compactly written as:
\begin{align} 
\boldsymbol{\mathcal H}_0 &= {\boldsymbol{\mathcal{F}}_0} {\boldsymbol{\mathcal{C}}_0} \boldsymbol{\mathcal\alpha}_0, \label{eq:H_0} \\
\boldsymbol{\mathcal H}_1 &= {\boldsymbol{\mathcal{F}}_1} {\boldsymbol{\mathcal{C}}_1} \boldsymbol{\mathcal\alpha}_1, \label{eq:H_1} \\
\boldsymbol{\mathcal H}_2^- &= {\boldsymbol{\mathcal{F}}_2} {\boldsymbol{\mathcal{C}}_2} {\boldsymbol{\mathcal{D}}_2} \boldsymbol{\mathcal\alpha}_2,  \label{eq:H_2-} \\
\boldsymbol{\mathcal H}_2^+ &= {\boldsymbol{\mathcal{F}}_2} {\boldsymbol{\mathcal{C}}_2} {\boldsymbol{\mathcal{D}}_2}^\text{H} \boldsymbol{\mathcal\alpha}_2,  \label{eq:H_2+} 
\end{align}
where ${\boldsymbol{\mathcal{F}}_i}$ are $(N \times L_i)$ matrices
whose $(n,l)$ entries are
\begin{equation} \label{eq:fourier_tau_all_i} 
\{{\boldsymbol{\mathcal{F}}_i}\}_{n,l} = \exp\left(-\jmath 2\pi \frac{n}{N}\tau_i^l \right), \quad i \in \{0,1,2\},
\end{equation}
with $n \in {\mathbb A}$ and $l = 0, \ldots, L_i-1$.
${\boldsymbol{\mathcal{C}}_i}$ are $(L_i \times L_i)$ diagonal matrices
whose diagonal entries are
\begin{equation} \label{eq:rotation_tau_all_i} 
\{{\boldsymbol{\mathcal{C}}_i}\}_{l,l} = \exp\left(-\jmath 2\pi \frac{F_\text{c}}{F_\text{samp}} \tau_i^l\right), \quad l = 0, \ldots, L_i-1.
\end{equation}
${\boldsymbol{\mathcal{D}}_2}$ is an $(L_2 \times L_2)$ diagonal matrix
whose entries are
\begin{equation} \label{eq:shift_tau_all_i} 
\{{\boldsymbol{\mathcal{D}}_2}\}_{l,l} = \exp\left(\jmath 2\pi \frac{F_\text{shift}}{F_\text{samp}} \tau_2^l\right), \quad l = 0, \ldots, L_2-1.
\end{equation}
$\boldsymbol{\mathcal\alpha}_i$ is an $(L_i \times 1)$ vector that contains the complex-valued channel coefficients of all $L_i$ paths
\begin{equation} \label{eq:alphas_i} 
\boldsymbol{\mathcal\alpha}_i=\big[\alpha_i^0, \ldots, \alpha_i^{L_i-1}\big]^\T.
\end{equation}

%% file: tag.tex
In this section we describe the tag's RF front-end operation on the incident illumination signal after it has passed through the TX-tag channel, the tag's data modulation, the frequency shifting operation, and then provide the expression of the frequency-shifted backscattered (reflected) signal.

\subsubsection{Tag RF front-end operation}
\label{sec:tag_RF_op}
The pass-band illumination signal $u(t)$ undergoes the TX-tag channel described by the impulse response $h_1(t)$ in \eqref{eq:multipath_ch01_discrete_freq_resp_tau_l}, and then is affected by the tag's circuitry twice: at incidence and at reflection. Let us denote the overall (two-way) tag's circuitry time-domain response by $g^\text{tag}(t)$. Therefore, the tag backscatters the signal
\begin{equation}
y(t) = u(t) \star h_1(t) \star g^\text{tag}(t). \label{eq:tag_passband_sig_conv}
\end{equation}
Using \eqref{eq:tx_analog_bb}--\eqref{eq:upconv_ofdm}, we obtain the frequency-domain representation of $y(t)$ in \eqref{eq:tag_passband_sig_conv} at the illuminator's subcarriers as an $N \times 1$ vector
\begin{equation}
\boldsymbol{\mathcal Y} = 
\boldsymbol{\mathcal S} \exp(\jmath\phi_{\text{TX}}) 
\odot \boldsymbol{\mathcal G}^\text{TX}
\odot \boldsymbol{\mathcal H}_1 
\odot \boldsymbol{\mathcal G}^\text{tag0},    
\label{eq:rx_vec} 
\end{equation}
where 
$\boldsymbol{\mathcal S}$ are the frequency domain illumination symbols given in \eqref{eq:ofdm_fd_vector},
$\phi_{\text{TX}}$ is the initial phase of the TX carrier wave in \eqref{eq:tx_carrier}, whereas
$\boldsymbol{\mathcal G}^\text{TX}$ and $\boldsymbol{\mathcal G}^\text{tag0}$ represent the TX's and tag's frequency responses at the illuminator subcarrier frequencies,
and $\boldsymbol{\mathcal H}_1$ is the TX-tag channel frequency response vector is given in \eqref{eq:multipath_ch01_discrete_freq_resp_tau_l} for $i=1$. 

The tag performs {\em passband reflection modulation} by controlling the reflection coefficient $\Gamma(t)$ of the RF circuit. The backscattered signal may be written as a multiplication of the illumination signal $y(t)$ and the signal generated locally, at the tag by changing the reflection coefficient $\Gamma(t)$, i.e.,
\begin{equation}
b(t) = y(t)[A_\text{s} - \Gamma(t)], \label{eq:backscattered_sig}
\end{equation} 
where $A_\text{s} \in {\mathbb C}$ represents {\em the structural mode scattering}, which is a time-invariable reflection component, attributed to the tag's geometry \cite{2010_BleDimSah}. The reflection coefficient $\Gamma(t)$, on the other hand, is variable and represents how the signal is reflected by the tag over time. The tag front-end can be electronically controlled to switch the tag antenna connection between two or more {\em loads}, resulting in modulation of the amplitude and/or phase of the backscattered signal. For example, by connecting the antenna to an open circuit, the reflection coefficient can take a value of $\Gamma_0 = 1$, which corresponds to full signal reflection, whereas a connection to a short circuit results in a value of $\Gamma_1 = -1 = e^{\jmath\pi}$, which is a full signal reflection with a phase reversal. This way, \ac{BPSK} modulation can be achieved. Throughout the literature there are several examples of higher-order modulation backscatter \cite{2012_ThWheTeiRey,2017_CorrBoaBor,2021_KimGeoDasTen}, but in this paper, the tag is assumed to perform BPSK modulation which requires the lowest implementation complexity.

\subsubsection{Tag Data Modulation Format}\label{sec:TagModulationFormat}
Let us assume that there are two distinct values $\Gamma_0,\Gamma_1 \in {\mathbb C}$ of the reflection coefficients corresponding to different backscatter modulation states. The data packet of $M$ modulating symbols generated at the tag is

\begin{equation} \label{eq:data_packet}
{\mathbf x} = [x_0,\ldots,x_{M-1}]^\T, \: x_m \in \{\Gamma_0,\Gamma_1\},
\: m \in \{0,\ldots,M-1\}.
\end{equation} 
We assume that the tag's modulation data symbol period is $T_\text{sym}$, thus the corresponding symbol rate is $F_\text{sym}=1/T_\text{sym}$. Since one data packet contains $M$ symbols,
the tag data packet duration is $T_\text{pkt}=MT_\text{sym}$.
The tag's baseband equivalent packet signal corresponding to a periodic sequence of $M$ symbols may be expressed as 
\begin{equation}
x_\text{pkt}(t) = \sum_{m=-\infty}^{\infty} x_{m \oplus M} \text{rect}_{[0,T_\text{sym})}(t-mT_\text{sym}). \label{eq:data_conv_details} 
\end{equation} 

The tag packet may consist of two parts: The first part includes a fixed preamble to detect the start of the packet and estimate the channel. The second part includes the tag ID and any other data to be communicated by the tag (payload). The design and evaluation of the tag packet is beyond the scope of this paper. 

\subsubsection{Frequency Shifting}
It is important to note that if the tag would modulate the multicarrier illumination signal $y(t)$ in \eqref{eq:tag_passband_sig_conv} directly with the rectangular data signal $x_\text{pkt}(t)$ in \eqref{eq:data_conv_details}, i.e., setting $\Gamma(t) = x_\text{pkt}(t)$ in \eqref{eq:backscattered_sig}, the corresponding backscattered signal $b(t)$ would interfere with the illumination signal $u(t)$ at the receiver after they pass through the channels $h_2(t)$ and $h_0(t)$, respectively.
That is because both the backscattered signal $b(t)$ and the illumination signal $y(t)$ would be centered at the same carrier frequency $F_\text{c}$.

In order to avoid the interference with the multicarrier illumination signal $y(t)$ centered at the carrier frequency $F_\text{c}$, the backscattered signal $b(t)$ is shifted to different frequencies, namely  the lower and upper sub-bands centered at $F_\text{c} \mp F_\text{shift}$.
This would be ideally achieved by multiplying the multicarrier illumination signal $y(t)$ by $\cos(2\pi F_\text{shift}t)$. However, the cosine function is complicated to generate in low-cost tag hardware, and a rectangular pulse sequence is used at the tag instead by switching the antenna impedance between open- and short-circuit at a fast rate. As a consequence, every harmonic component of the rectangular clock signal will be modulated by the multicarrier illumination signal\footnote{The even harmonics of the rectangular clock signal (including the DC component) are null, hence, only odd harmonics will be modulated.}.

\begin{figure}[!t]
\centering
\includegraphics[width=\columnwidth]{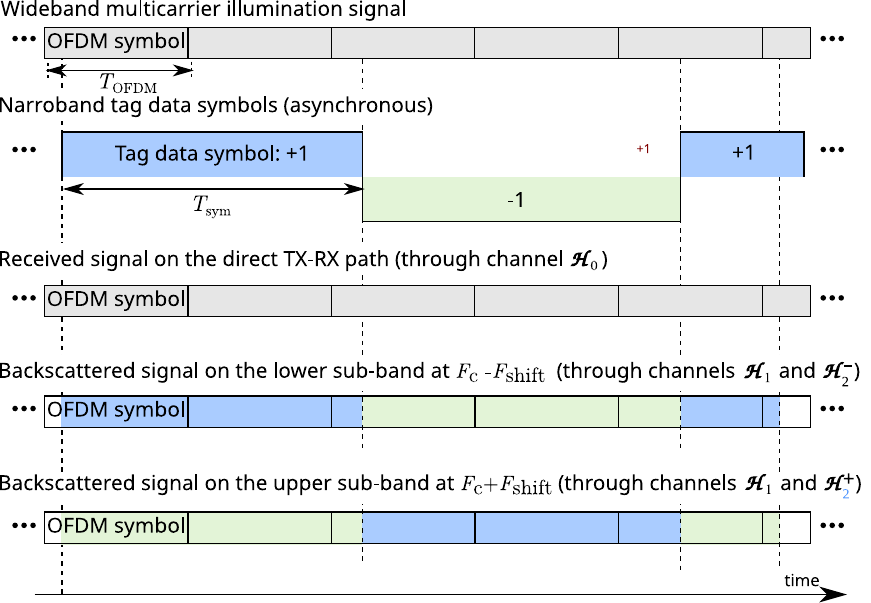}
\caption{The multicarrier illumination signal, the tag data signal, and the received signals at the output of the direct TX-RX channel $\boldsymbol{\mathcal H}_0$, as well as the received backscattered signal on the lower and upper sub-bands $\boldsymbol{\mathcal H}_2^-$ and $\boldsymbol{\mathcal H}_2^+$, respectively. \label{fig:backscatter_multicarrier_illumination_and_tag_data} }
\end{figure}

We define the ``clock-like" bipolar rectangular signal 
\begin{equation} \label{eq:clock_periodic}
\kappa(t) = \sum_{m=-\infty}^{\infty} (-1)^m~\text{rect}_{[0,T_\text{clk})}(t - mT_\text{clk}),
\end{equation}
where $T_\text{clk}$ is the clock "half-period" corresponding to either state $+1$ or $-1$.
The clock $\pm  1$-states change $2Q$ times faster than the data symbol's rate, i.e., with frequency 
\begin{equation} \label{eq:F_shift}
F_\text{shift} = Q F_\text{sym}.
\end{equation}
Frequency shifting is performed by multiplying the rectangular data signal $x_\text{pkt}(t)$ in \eqref{eq:data_conv_details} by the clock signal $\kappa(t)$ in \eqref{eq:clock_periodic}. Therefore, the frequency-shifted backscattered signal is obtained by changing the reflection coefficient according to
\begin{equation} \label{eq:frequency_shifted_signal}
\Gamma(t) \triangleq x_\text{pkt}(t)\kappa(t).
\end{equation}
The amplitude of the harmonics of the rectangular clock signal decrease proportionally to their order, but we are interested in particular in the first-order harmonics located at $F=\mp F_\text{shift}$, which correspond to the lower and upper sub-bands of interest,\footnote{The higher order harmonics (third, fifth and so on), are attenuated by the band-pass characteristics of the tag, and also heavily filtered at the RX, hence we ignore them in the model.}
as illustrated in Figure~\ref{fig:channel_spectrum}. 
The first harmonics of the baseband-equivalent clock signal in \eqref{eq:clock_periodic} have complex amplitudes 
\begin{equation} 
{\mathcal K}(\mp F_\text{shift}) = \pm \frac{2\jmath}{\pi}, \label{eq:kappa_1st_spectral_lines} 
\end{equation} 
but since backscatter modulation takes place in pass-band, the first harmonics  of the multicarrier illumination signal will be shifted to $F_\text{c}\mp F_\text{shift}$ where the lower and upper bands are centered, respectively. In the central band, the tag only reflects due to the structural mode scattering $A_\text{s}$, as seen in \eqref{eq:backscattered_sig}. 
As shown in \eqref{eq:frequency_shifted_signal}, the illumination signal in the lower and upper band, respectively, will be modulated by the clock signal $\kappa(t)$ in \eqref{eq:clock_periodic}, as well as by the tag data $x_\text{pkt}(t)$ in \eqref{eq:data_conv_details}. 
The duration of the OFDM symbol, denoted as $T_\text{OFDM}$ is much smaller than the duration of the tag data symbol $T_\text{sym}$, as shown in Figure~\ref{fig:backscatter_multicarrier_illumination_and_tag_data}. 
Consequently, within the time window corresponding to one OFDM symbol, the signal $x_\text{pkt}(t)$ appears to be quasi-static, taking the of value the the instantaneous tag data symbol $x_m$ at the given time instance $t_m$.
Considering the illumination signal spectrum in \eqref{eq:rx_vec}
and the upper/lower fist harmonic of the clock in \eqref{eq:kappa_1st_spectral_lines},
the backscattered signals corresponding to the central, lower and upper sub-band subcarrier indices $n \in {\mathbb A}$, respectively, can be represented at time $t_m$ by the following $N \times 1$ vectors:
\begin{align}
\boldsymbol{\mathcal B}^0 (t_m) 
&= \frac{1}{2}A_\text{s} \exp(\jmath\phi_{\text{TX}})
\boldsymbol{\mathcal H}_1 \odot \boldsymbol{\mathcal G}^\text{TX} \odot \boldsymbol{\mathcal G}^\text{tag0} \odot \boldsymbol{\mathcal S}, \label{eq:B0_final} \\
\boldsymbol{\mathcal B}^\mp (t_m) 
&= \mp\frac{\jmath}{\pi}x_m\exp(\jmath\phi_\text{TX})
\boldsymbol{\mathcal H}_1 \odot \boldsymbol{\mathcal G}^\text{TX} \odot \boldsymbol{\mathcal G}^\text{tag$\mp$} \odot \boldsymbol{\mathcal S}, \label{eq:B-+_final}
\end{align}
where 
and $\boldsymbol{\mathcal G}^\text{tag$\mp$}$ represent the tag's frequency response at the subcarriers corresponding to the 
lower and upper bands, respectively. 
Since the illumination signal is not synchronized in any way with the tag's modulation signal, the symbol boundaries may fall anywhere within the window of the OFDM symbol, as illustrated in Fig. \ref{fig:backscatter_multicarrier_illumination_and_tag_data}. 
The tags' data symbol demodulation will be addressed later in Section~\ref{sec:tag_symbols}.

%% file: reader.tex
\begin{figure}
\centering
\includegraphics[width=\columnwidth]{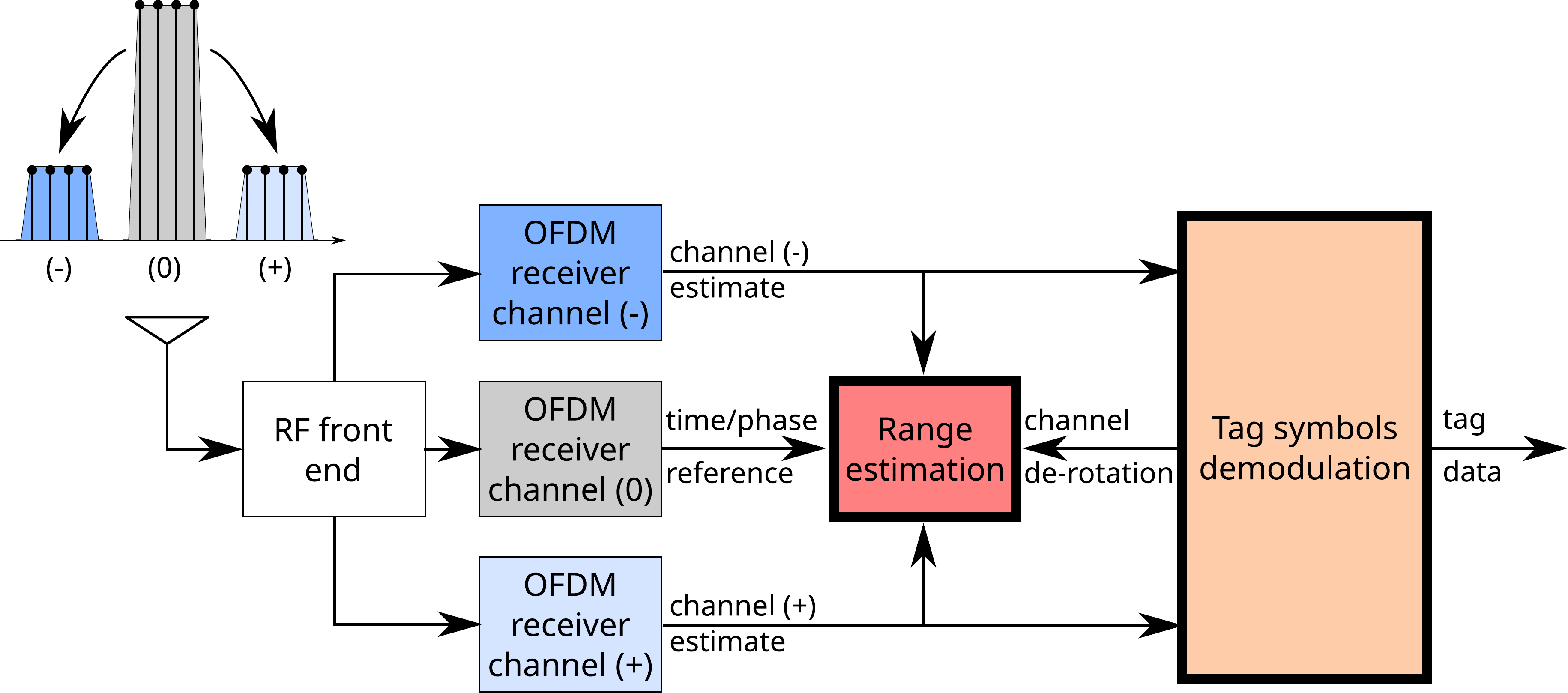}
\caption{The high-level \ac{OFDM}-based reader (RX) block diagram.
The stronger TX-RX signal in the central band is used for time/phase synchronization, which is then used for the backscatter signal in the lower and upper bands.
This enable reliable ranging and symbol demodulation.
\label{fig:RX_implementation_high-level} }
\end{figure}

The triple-band receiver processing blocks are shown in \figref{fig:RX_implementation_high-level}. 
The stronger TX-RX signal in the central band is used to reliably retrieve the time/phase synchronization information, which is then used to process the backscatter signal in the lower and upper bands. This enables robust ranging and symbol demodulation.
The processing takes place in a per-band fashion, i.e., for the central, lower, and upper bands, respectively. In our system, we assume that the corresponding local oscillators used for down-conversion of all bands are derived from the same clock. We also assume that the \ac{ADC} sampling is simultaneous on the three channels (i.e. sampling is triggered at the same time).

\subsubsection{Band-pass filtering and amplification}\label{sec:Rx-BPF}

Additional frequency-dependent attenuation and delay (hence also phase shift) is experienced by the signal while passing through the RX analog front-ends, as well as other circuitry and connection feeds, digital filters and processing.
The overall transceiver circuitry response can be modeled as a frequency response vector of complex gains in each band.
This can be regarded as a static ``channel" whose characteristics can be determined accurately by calibration (e.g., by taking over-the-air measurements at known distances in anechoic chamber). Let us denote the overall RX circuitry (from the antenna to the \ac{ADC}) baseband equivalent time-domain responses by $g^\text{RX0}(t)$, $g^\text{RX$\mp$}(t)$ for the central, lower and upper band respectively, and their corresponding frequency responses at the subcarrier frequencies of each band to be characterized by the $N \times 1$ vectors $\boldsymbol{\mathcal G}^\text{RX0}$, $\boldsymbol{\mathcal G}^\text{RX$\mp$}$, respectively.

The received signals after the bandpass filtering and low noise amplifier (LNA) on the central, lower and upper band RX branches can be derived assuming bandpass filters whose pass-band is $B$ and whose stop-band attenuation is sufficiently high, such that the contribution of adjacent bands to the band of interest is negligible. 
The band-pass filtered and amplified signal on the central band consists of two terms:
the first high-power term corresponds to the direct TX-RX channel transmission. The second term, whose power is typically several tens of decibels lower compared to the first term, corresponds to the structural mode scattering (See Sec.~\ref{sec:tag_RF_op}). Since this term has significantly lower magnitude, it can be incorporated in the noise plus interference term for this analysis.
Therefore, the illumination signal $u(t)$ in \eqref{eq:upconv_ofdm}, after it passes through the 
TX circuitry characterized by $g^\text{TX}(t)$, 
the TX-RX channel $h_0(t)$ in \eqref{eq:multipath_ch0_imp_resp}, and the RX central band circuitry characterized by $g^\text{RX0}(t)$, can be called the bandpass received signal for the central band, and can be defined as  
\begin{equation}
r_\text{BP}^0(t) = u(t) \star g^\text{TX}(t)\star h_0(t) \star g^\text{RX0}(t) + v_\text{BP}^0(t),
\label{eq:rx_signal0} 
\end{equation}
where $v_\text{BP}^0(t)$ consists of contributions of the thermal noise, residual inter-band interferences, structural scattering, and other disturbances.
We assume that $v_\text{BP}^0(t)$ is zero-mean with variance $(\sigma_v^0)^2$, and that its Fourier transform is ${\mathcal V}_\text{BP}^0(F)$.

The received signals corresponding to the lower and upper bands are obtained by passing the backscattered signal $b(t)$ in \eqref{eq:backscattered_sig} through the TX-tag channel $h_2(t)$ in \eqref{eq:multipath_ch0_imp_resp} (for $i=2$), and then through the lower and upper band RX front-ends characterized by the equivalent responses $g^\text{RX$-$}(t)$ and $g^\text{RX$+$}(t)$, respectively.
Again, we may drop the structural mode term because it falls in the central band and is rejected by the analog filtering. Therefore, we obtain
\begin{equation}
r_\text{BP}^\mp(t) = b(t) \star h_2(t) \star g^{\text{RX}\mp}(t)  + v_\text{BP}^\mp(t),
\label{eq:rx_signal-+}
\end{equation}
where
$v_\text{BP}^-(t)$ and $v_\text{BP}^+(t)$ are the noise plus interference terms corresponding to the lower and upper bands, respectively. They consist of contributions of the thermal noise, residual inter-band interferences, and other disturbances. We assume that $v_\text{BP}^-(t)$ and $v_\text{BP}^+(t)$ are zero-mean noises with variances $(\sigma_v^-)^2$ and $(\sigma_v^+)^2$, respectively. 

\subsubsection{Down-conversion and low-pass filtering}

We assume that the down-conversion carrier waveforms  at the RX for each of the bands are derived from the same clock, hence they experience common \ac{CFO} $F_\text{off}^\text{RX}$, and initial phases $\phi_{\text{RX}}, \phi_{\text{RX}}^\mp$, respectively. For the central, lower and upper band, the down-conversion carrier are:
\begin{align}
c^\text{RX0}(t) =& \exp \left\{ -\jmath [2\pi (F_\text{c} + F_\text{off}^\text{RX})t + \phi_{\text{RX}}] \right\},  \label{eq:rx_carrier0} \\
c^\text{RX$\mp$}(t) =& \exp \left\{ -\jmath [2\pi (F_\text{c} \mp F_\text{shift} + F_\text{off}^\text{RX$\mp$})t + \phi_{\text{RX}}^\mp] \right\},  \label{eq:rx_carrier-+} 
\end{align}
where $F_\text{off}^\text{RX}$ is the common RX carrier frequency offset of the central band carrier w.r.t. TX carrier, and $F_\text{off}^\text{RX,tag$\mp$}$
is the combined RX \ac{CFO} and tag \ac{CFO} on the lower and upper bands, respectively 
\begin{equation}
F_\text{off}^\text{RX,tag$\mp$} = F_\text{off}^\text{RX} \mp F_\text{off}^\text{tag}, \\
\end{equation}
and the tag \ac{CFO} $F_\text{off}^\text{tag}$ is caused by imperfect frequency shifting $F_\text{shift}$ corresponding to the tag's clock $\kappa(t)$ in \eqref{eq:clock_periodic}.
Down-conversion is achieved by multiplying the band-pass received signals on the central, lower and upper bands in \eqref{eq:rx_signal0}--\eqref{eq:rx_signal-+}, with the RX carrier waves in 
\eqref{eq:rx_carrier0}--\eqref{eq:rx_carrier-+}, respectively
\begin{align}
r_\text{bb}^0(t) &= r_\text{BP}^0(t)c^\text{RX}(t), \label{eq:rx_bb_0} \\
r_\text{bb}^\mp(t) &= r_\text{BP}^\mp(t)c^{\text{RX}\mp}(t),   \label{eq:rx_bb_-+}
\end{align}
Then, the baseband signals in each band are obtained by low-pass filtering.
Using \eqref{eq:rx_signal0}--\eqref{eq:rx_carrier-+}, \eqref{eq:tag_passband_sig_conv} and \eqref{eq:backscattered_sig},
we obtain 
\begin{align}
r_\text{bb}^0(t) =& \frac{1}{2} \exp \left[ \jmath (+2\pi F_\text{off}^\text{RX}t + \phi_\text{TX} - \phi_\text{RX}) \right] \cdot \nonumber \\
& \cdot [s(t) \star g^0(t)\star h_0(t)] + v_\text{bb}^0(t), \label{eq:rx_bb_0_detailed} \\
r_\text{bb}^\mp(t) =& \mp\frac{\jmath}{\pi} \exp \left[+\jmath (2\pi  F_\text{off}^\text{RX,tag$\mp$} t + \phi_{\text{RX}}^\mp-\phi_{\text{TX}}) \right] \cdot \nonumber \\
& \cdot [s(t) \star g^\mp(t)  \star h_1(t) \star h_2(t) ] x_\text{pkt}(t) + v_\text{bb}^\mp(t),
\label{eq:rx_bb_-+_detailed}
\end{align}
where for brevity, we denoted the overall TX-RX circuitry impulse response by 
$g^0(t) = g^\text{TX}(t) \star g^\text{RX0}(t)$,
and the TX-tag-RX circuitry impulse responses by
$g^\mp(t) = g^\text{TX}(t) \star g^\text{tag}(t) \star g^{\text{RX}\mp}(t)$,
and
$v_\text{bb}^0(t) = v_\text{BP}^0(t) c^\text{RX0}(t)$,
$v_\text{bb}^-(t) = v_\text{BP}^-(t) c^\text{RX$-$}(t)$, and 
$v_\text{bb}^+(t) = v_\text{BP}^+(t) c^\text{RX$+$}(t)$ 
are the down-converted and low pass-filtered noise terms in the central lower and upper bands,
respectively, and 
${\mathcal V}_\text{bb}^0(F)$,
${\mathcal V}_\text{bb}^-(F)$, and 
${\mathcal V}_\text{bb}^+(F)$ are their Fourier transforms.
The frequency-domain descriptions within one \ac{OFDM} symbol window of the baseband signal on the central band $r_\text{bb}^0(t)$ in \eqref{eq:rx_bb_0}, and the baseband signal on the lower and upper bands $r_\text{bb}^-(t)$ and $r_\text{bb}^+(t)$ in \eqref{eq:rx_bb_-+}, respectively, at time $t_m$ corresponding to the $m$th tag data symbol $x_m$ are
\begin{align}
{\mathcal R}_\text{bb}^0(F,t_m) &= \frac{1}{2}\exp[\jmath(\phi_{\text{RX}}-\phi_{\text{TX}})] {\mathcal H}_0(F) \cdot \nonumber \\ 
& \cdot \sum_{n=-N/2}^{N/2-1} {\mathcal S}_n {\mathcal G}_n^0 \delta[F - (F_\text{off}^\text{RX} + n\Delta F)] + {\mathcal V}_\text{bb}^0(F), \\
{\mathcal R}_\text{bb}^\mp(F,t_m) & =\mp\frac{\jmath}{\pi}\exp[\jmath(\phi_{\text{RX}}^\mp-\phi_{\text{TX}})] x_m 
{\mathcal H}_1(F){\mathcal H}_2(F) \cdot \nonumber \\ 
& \cdot
\sum_{n=-N/2}^{N/2-1} {\mathcal S}_n {\mathcal G}_n^\mp(F) \delta\left[F - (F_\text{off}^\text{RX,tag$\mp$}  + n\Delta F) \right] +  \nonumber \\ 
&+ {\mathcal V}_\text{bb}^\mp(F),
\end{align}
where $x_m$ is the instantaneous tag data symbol value at time $t_m$ of the tag data signal $x_\text{pkt}(t)$ in \eqref{eq:data_conv_details}.
\subsubsection{Analog-to-digital conversion}
Let us assume that the \ac{ADC} samples the signal with the sampling period $T_\text{samp}$. Hence, assuming 
\begin{equation}
t = k T_\text{samp}, \quad k \in {\mathbb Z},  \label{eq:time_discretization} 
\end{equation}
all continuous time signals can be discretized in time by replacing the continuous time $t$ with a discrete sample index $k$. Similarly, the frequency $F$ will be written in terms of normalized frequency 
\begin{equation}
f = \frac{F}{F_\text{samp}}, \quad -\frac{1}{2} \leq f < \frac{1}{2}  \label{eq:normalized_freq}
\end{equation}
and all frequency notations that have subscripts are rewritten in terms of the corresponding normalized frequencies whenever needed (for example, $F_\text{off}^\text{RX}$ appears in the discrete time domain as $f_\text{off}^\text{RX}= F_\text{off}^\text{RX}/F_\text{samp}$).
From now on, all the processing takes place in the digital domain. 
In addition, the time-discrete signals in the central, lower and band undergo amplitude discretization, hence are affected by additional uniformly distributed quantization noises $v_\text{q}^0$, $v_\text{q}^-$, $v_\text{q}^+$, respectively.

\subsubsection{Overall CFO correction\label{sec:cfo}}
Let us assume that \ac{CFO} estimation is done in a per-band basis using prior art methods. We denote by $\hat{f}_\text{off}^\text{RX}$ the central band's coarse \ac{CFO} estimate of the true \ac{CFO} between TX and RX, $f_\text{off}^\text{RX}$.
Similarly, we denote by $\hat{f}_\text{off}^\text{RX$\mp$,tag}$ the lower/upper band estimates of the true combined TX-tag-RX \ac{CFO} $f_\text{off}^\text{RX$\mp$, tag} = f_\text{off}^\text{RX} \mp f_\text{off}^\text{tag}$ on the lower and upper band, respectively. 
In the presence of noise and imperfections, the \ac{CFO} estimates will deviate from the true values, and some residual \ac{CFO} value will still remain in each band. Let us define the residual \ac{CFO}s after \ac{CFO} correction for the central band as $f_\epsilon^0$, 
and for the upper and lower band as $f_\epsilon^\mp$ as follows:
\begin{eqnarray} 
f_\epsilon^0 & =& \hat{f}_\text{off}^\text{RX} - f_\text{off}^\text{RX} \label{eq:cfo_res_0} \\
f_\epsilon^\mp & =& \hat{f}_\text{off}^\text{RX$\mp$,tag} - f_\text{off}^\text{RX$\mp$,tag}. \label{eq:cfo_res-+} 
\end{eqnarray}
Based on \eqref{eq:time_discretization}--\eqref{eq:normalized_freq}, the discretized baseband signals in \eqref{eq:rx_bb_0_detailed}--\eqref{eq:rx_bb_-+_detailed} can be given by
\begin{align}
r_\epsilon^0(k)
=& \frac{1}{2} \exp \left[ \jmath (+2\pi f_\epsilon^0k + \phi_\text{TX} - \phi_\text{RX}) \right] \cdot \nonumber \\
&\cdot [s(k) \star h_0(k) \star g^0(k)] + v_\epsilon^0(k) \label{eq:rx_bb_0_samp_coarse_cfo_corr}  \\
r_\epsilon^\mp(k)
=&\mp\frac{\jmath}{\pi} \exp \left[+\jmath (2\pi  f_\epsilon^\mp k + \phi_{\text{RX}}^\mp-\phi_{\text{TX}}) \right]  \cdot \nonumber \\
&\cdot [s(k)   \star h_{1,2}(k) \star g^\mp(k)] x_\text{pkt}(k) + v_\epsilon^\mp(k) \label{eq:rx_bb_-+_samp_coarse_cfo_corr} 
\end{align}
where $v_\epsilon^0(k) = v_\text{q}^0 + v_\text{bb}^0(k)\exp(-\jmath 2 \pi \hat{f}_\text{off}^\text{RX})$, and 
$v_\epsilon^\mp(k) = v_\text{bb}^\mp(k)\exp(-\jmath 2 \pi k \hat{f}_\text{off}^\text{RX$\mp$,tag})$
are the rotated sampled noise after CFO correction in the central lower and upper bands, respectively, that also include the quantization noise to due \ac{ADC}.

The frequency-domain descriptions of the digitized signal after \ac{CFO} correction on the central band $r_\epsilon^0(k)$ in \eqref{eq:rx_bb_0_samp_coarse_cfo_corr}, and on the lower and upper bands $r_\epsilon^\mp(k)$ in \eqref{eq:rx_bb_-+_samp_coarse_cfo_corr}, respectively, within one \ac{OFDM} symbol window at time $t_m$ corresponding to the $m$th tag data symbol $x_m$ are:
\begin{align}
{\mathcal R}_\epsilon^0(f,t_m)
=& \frac{1}{2}\exp[\jmath(\phi_{\text{RX}}-\phi_{\text{TX}})] {\mathcal H}_0(f)  \cdot \nonumber \\
&\cdot \sum_{n=-N/2}^{N/2-1} {\mathcal S}_n {\mathcal G}_n^0 \delta\left[f - \left(f_\epsilon^0 + \frac{n}{N}\right)\right] + {\mathcal V}_\epsilon^0(f), \label{eq:rx_bb_0_spec_samp_coarse_cfo_corr} \\
{\mathcal R}_\epsilon^\mp(f,t_m) 
=& \mp\frac{\jmath}{\pi}\exp[\jmath(\phi_{\text{RX}}^\mp-\phi_{\text{TX}})]x_m {\mathcal H}_1(f) {\mathcal H}_2(f)
\cdot \nonumber \\
&\cdot \sum_{n=-N/2}^{N/2-1} {\mathcal S}_n {\mathcal G}_n^\mp(f) \delta\left[f - \left(f_\epsilon^\mp  + \frac{n}{N}\right)\right] + {\mathcal V}_\epsilon^\mp(f), \label{eq:rx_bb-spec_samp_coarse_cfo_corr}
\end{align}
where ${\mathcal V}_\epsilon^0(f)$, ${\mathcal V}_\epsilon^\mp(f)$ are the  Fourier transforms of the time-domain noise samples $v_\epsilon^0(f)$, $v_\epsilon^\mp(f)$, respectively.
\subsubsection{\ac{OFDM} symbol timing}
\ac{OFDM} symbol timing can be recovered by correlating the received time-domain signal with a known transmitted preamble sequence $s(k)$ with excellent correlation properties (for example, a Zadoff-Chu sequence). The time-synchronized \ac{OFDM} symbol is obtained at the output of a sliding correlator when the corresponding correlation function reaches its maximum absolute value. Considering the simultaneous sampling of the three bands, reliable timing estimation can be obtained from the central band whose SNR is substantially higher. Then this timing correction can be applied to all bands. This option is expected to improve the range estimation, and also reduces the complexity of the receiver because only one sliding correlator is needed. However, this option requires a more careful bookkeeping of the \ac{OFDM} symbol edges due to the different calibration delays in each RX branch. Then the timing estimate is obtained for each block of $N$ samples entering the three sliding correlators as the index maximizing the correlation absolute value with the known preamble $s(k)$:
\begin{equation}
k_0 = \arg\max_k \left|\sum_{n=0}^{N-1} s^*(k) r_\epsilon^0(k-n))\right|.
\label{eq:ind_corr_max}
\end{equation}
The estimated \ac{OFDM} symbol time-domain samples are obtained from 
\eqref{eq:rx_bb_0_samp_coarse_cfo_corr}, \eqref{eq:rx_bb_-+_samp_coarse_cfo_corr}, and \eqref{eq:ind_corr_max} as
\begin{align}
\hat{\mathbf r}_\epsilon^0(k_0) &= [r_\epsilon^0(k_0),\ldots,r_\epsilon^0(k_0+N-1)]^\T, \label{eq:ofdm_sym_0} \\
\hat{\mathbf r}_\epsilon^\mp(k_0) &= [r_\epsilon^\mp(k_0),\ldots,r_\epsilon^\mp(k_0+N-1)]^\T,  \label{eq:ofdm_sym_-+} 
\end{align}
The timing estimate will contain errors due to unresolvable multipath, noise and other imperfections. Assuming the normalized timing error (i.e., the fractional timing error in samples) for the central band is $\hat\zeta^0$, this will be reflected in frequency domain as linear phase rotations across the OFDM subcarriers. This corresponds to multiplying the the subcarriers by an $N \times N$ diagonal matrix whose diagonal entries are
\begin{equation} 
\{\boldsymbol{\mathcal D}_\zeta^0\}_{n,n} =
\exp\left(-\jmath 2\pi \frac{n}{N} \hat\zeta^0 \right),
\quad n \in {\mathbb A}. \label{eq:D_zeta_0} 
\end{equation}
\subsubsection{\ac{FFT} operation at RX}
The estimated OFDM symbols in frequency domain can be obtained by taking an $N$-point \ac{FFT} on the vectors of samples corresponding to each channel given in~\eqref{eq:ofdm_sym_0}--\eqref{eq:ofdm_sym_-+}:
\begin{align}
\hat{\boldsymbol{\mathcal R}}_\epsilon^0(k_0) 
&= {\mathbf F}_N \hat{\mathbf r}_\epsilon^0(k_0)
= \frac{1}{2}\exp[\jmath(\phi_{\text{RX}}-\phi_{\text{TX}})] \nonumber \\
& \boldsymbol{\mathcal D}_\zeta^0 \boldsymbol{\mathcal D}_\epsilon^0 \cdot [\boldsymbol{\mathcal H}_0 \odot \boldsymbol{\mathcal G}^0 \odot \boldsymbol{\mathcal S}] + \boldsymbol{\mathcal V}_\epsilon^0, \label{eq:rx_ofdm_0_spec} \\
\hat{\boldsymbol{\mathcal R}}_\epsilon^\mp (k_0)
&= {\mathbf F}_N \hat{\mathbf r}_\epsilon^\mp(k_0)
= \mp\frac{\jmath}{\pi}\exp[\jmath(\phi_{\text{RX}}-\phi_{\text{TX}})] x_\text{pkt}(k_0)
 \nonumber \\
&\boldsymbol{\mathcal D}_\zeta^0 \boldsymbol{\mathcal D}_\epsilon^\mp \cdot [\boldsymbol{\mathcal H}_1 \odot \boldsymbol{\mathcal H}_2^\mp \odot \boldsymbol{\mathcal G}^\mp \odot \boldsymbol{\mathcal S}]
+ \boldsymbol{\mathcal V}_\epsilon^\mp, \label{eq:rx_ofdm+-spec}
\end{align}
where
$\boldsymbol{\mathcal V}_\epsilon^0,\boldsymbol{\mathcal V}_\epsilon^-$, and 
$\boldsymbol{\mathcal V}_\epsilon^+$ are the $(N \times 1)$ frequency-domain noise terms corresponding to the central, lower, and upper band, respectively.
$\boldsymbol{\mathcal G}^0,\boldsymbol{\mathcal G}^-,\boldsymbol{\mathcal G}^+$ are $(N \times 1)$ overall gain vectors containing the complex gains
at the $n$th subcarrier frequencies in the central, lower and upper band, respectively.
Matrix $\boldsymbol{\mathcal D}_\zeta^0$ is defined in \eqref{eq:D_zeta_0}.
Matrices $\boldsymbol{\mathcal D}_\epsilon^0$, $\boldsymbol{\mathcal D}_\epsilon^-$ and $\boldsymbol{\mathcal D}_\epsilon^+$
in \eqref{eq:rx_ofdm_0_spec}--\eqref{eq:rx_ofdm+-spec}, respectively,
are full $(N \times N)$ matrices expressing the inter-carrier interference (ICI) caused by the residual \ac{CFO}s errors
$f_\epsilon^0$ and $f_\epsilon^\mp$, on the central and lower/upper bands, respectively. Their $(n_1,n_2)$-entries are given by \cite{2013_AbrHagKoi}

\begin{align}  
\{\boldsymbol{\mathcal D}_\epsilon^0\}_{n_1,n_2} &= 
\exp\left[-\jmath\frac{\pi}{N}(n_1-n_2+Nf_\epsilon^0)\right] \cdot \nonumber \\
&\cdot \frac{\sin\left[\pi(n_1-n_2+Nf_\epsilon^0)\right]}{N\sin\left[\frac{\pi}{N}(n_1-n_2+Nf_\epsilon^0)\right]}, \label{eq:D_epsilon_0}
\\
\{\boldsymbol{\mathcal D}_\epsilon^\mp\}_{n_1,n_2} &=
\exp\left[-\jmath\frac{\pi}{N}(n_1-n_2+Nf_\epsilon^\mp)\right] \cdot \nonumber \\
&\cdot \frac{\sin\left[\pi(n_1-n_2+Nf_\epsilon^\mp)\right]}{N\sin\left[\frac{\pi}{N}(n_1-n_2+Nf_\epsilon^\mp)\right]},\quad n_1,n_2 \in {\mathbb A}.  \label{eq:D_epsilon+-}
\end{align}

\subsubsection{Channel\protect\footnote{The term {\em ``channel"} here does not refer to just contributions of the propagation environment, but also additional group delay because of the receive chain, filtering, amplification, etc in each band.} estimation}
The channel input is the vector of transmitted frequency-domain \ac{OFDM} symbols at the illuminator $\boldsymbol{\mathcal S}$, and its three outputs are the received \ac{OFDM} symbol vectors after \ac{CFO} correction and timing in each band given in \eqref{eq:rx_ofdm_0_spec}--\eqref{eq:rx_ofdm+-spec}.
This means that, for example for the central band, apart from the TX-RX propagation channel $\boldsymbol{\mathcal H}_0$, the estimated channel also includes the TX and RX combined gain $\boldsymbol{\mathcal G}^0$, the carrier phase difference between RX and TX $\phi_{\text{RX}}-\phi_{\text{TX}}$, the residual \ac{CFO} $f_\epsilon^0$, the timing error $\zeta^0$, and all the corresponding imperfections and noise $\boldsymbol{\mathcal V}_\epsilon^0$, as shown in \eqref{eq:rx_ofdm_0_spec}.
Similarly, for the lower and upper bands, apart from the TX-tag propagation channel $\boldsymbol{\mathcal H}_1$ and tag-RX propagation channel $\boldsymbol{\mathcal H}_2^-,\boldsymbol{\mathcal H}_2^+$, the estimated channels also include the TX, tag and RX combined gains $\boldsymbol{\mathcal G}^-,\boldsymbol{\mathcal G}^+$, the carrier phase difference between RX and TX $\phi_{\text{RX}}^\mp-\phi_{\text{TX}}$, the remaining \ac{CFO}s $f_\epsilon^-,f_\epsilon^+$, the timing error $\zeta^0$, and all the corresponding imperfections and noise
$\boldsymbol{\mathcal V}_\epsilon^-,\boldsymbol{\mathcal V}_\epsilon^+$, as shown in \eqref{eq:rx_ofdm+-spec}. 
We assume that the \ac{CFO} on each band have been corrected, and the residual \ac{CFO}s $f_\epsilon^0,f_\epsilon^-,f_\epsilon^+$ are negligible compared to subcarrier spacing. In that case, matrices $\boldsymbol{\mathcal D}_\epsilon^0$, $\boldsymbol{\mathcal D}_\epsilon^-$ and $\boldsymbol{\mathcal D}_\epsilon^+$ can be approximated by the identity matrices.  

Frequency-domain channel estimation is then performed in an OFDM-grid basis. The complex gains of the overall channels corresponding to each subcarrier are obtained by element-wise division of the received frequency-domain symbol after \ac{CFO} compensation and timing in each band by the transmitted frequency-domain symbol $\boldsymbol{\mathcal S}$. Using \eqref{eq:rx_ofdm_0_spec}--\eqref{eq:rx_ofdm+-spec}, we obtain
\begin{align}
\hat{\boldsymbol{\mathcal H}}^0(k_0)  =&
\hat{\boldsymbol{\mathcal R}}_\epsilon^0 \oslash \boldsymbol{\mathcal S}
= \frac{1}{2}\exp[\jmath(\phi_{\text{RX}}-\phi_{\text{TX}})]\nonumber\\
&\cdot\boldsymbol{\mathcal D}_\zeta^0 
\diag{\boldsymbol{\mathcal G}^0} 
\boldsymbol{\mathcal H}_0 
+ \boldsymbol{\mathcal V}^0,
\label{eq:ch_est_0_spec}
\\
\hat{\boldsymbol{\mathcal H}}^\mp(k_0) =&
\hat{\boldsymbol{\mathcal R}}_\epsilon^\mp \oslash \boldsymbol{\mathcal S}
= \mp\frac{\jmath}{\pi}\exp[\jmath(\phi_{\text{RX}}^\mp-\phi_{\text{TX}})] x_\text{pkt}(k_0)\cdot \nonumber \\
&\cdot 
\boldsymbol{\mathcal D}_\zeta^0 \diag{\boldsymbol{\mathcal G}^\mp} 
(\boldsymbol{\mathcal H}_1 \odot \boldsymbol{\mathcal H}_2^\mp) 
+ \boldsymbol{\mathcal V}^\mp, \label{eq:ch_est+-spec}
\end{align}
where the new noise terms in the central, lower and upper band are
$\boldsymbol{\mathcal V}^0 = \boldsymbol{\mathcal V}_\epsilon^0 \oslash \boldsymbol{\mathcal S}$,
$\boldsymbol{\mathcal V}^\mp = \boldsymbol{\mathcal V}_\epsilon^\mp \oslash \boldsymbol{\mathcal S}$, respectively.

\subsubsection{Tag signal extraction\label{sec:tag_symbols}}

In \eqref{eq:ch_est+-spec}, we may notice the dependence of the \ac{OFDM} channel estimates on the tag's data symbols $x_\text{pkt}(k_0)$, e.g. a data symbol may or may not ``flip" the channel estimate, if the tag is performing binary phase modulation. 
This means that for range estimation, the lower and upper band channel estimates corresponding to the TX-tag-RX propagation environment need to be de-rotated by the tag symbols, to cancel out the effect of modulation on the channel estimate. 
The de-rotation requires estimating the tag symbols $x_\text{pkt}(k_0)$, which we brief in this section.

Note that the duration of the tag data symbol is at least a few times larger than the duration of the illuminator's \ac{OFDM} symbol. Therefore, the tag symbols are oversampled and each sample is associated to one OFDM symbol. Accordingly, all sub-carriers of the OFDM symbol can be used to retrieve the tag sample $x_\text{pkt}(k_0)$. 

The tag sample estimation consists of three steps: \textit{(a)} identify the beginning of a tag packet based on the preamble correlation, \textit{(b) }estimate the effective channel observed during the tag backscattering by using the preamble of the tag packet, and finally, \textit{(c)} demodulating the tag samples. 

The beginning of the packet can be identified using the correlation with a known preamble. Notably, a reliable packet detection also requires CFO estimation $f_\epsilon^-,f_\epsilon^+$ on the lower and upper bands and accurate tracking of the CFO and phase over the entire tag packet. To estimate and track CFO and phase, many existing techniques such as digital \ac{PLL} and decision-directed phase tracking can be used. The performance analysis of these methods is beyond the scope of this work. 

Let $k_\text{pkt}$ denote the index of the first sample of an OFDM symbol associated to the beginning of the tag packet. 
Assuming the tag packet starts with a preamble, $x_\text{pkt}(k_\text{pkt})$ is known to the receiver. Hence, we can use the known preamble $x_\text{pkt}(k_\text{pkt})$ to estimate the channel observed by the tag packet as follows: 
\begin{align}
   \hat{\boldsymbol{\mathcal H}}_\text{pkt}^\mp = \hat{\boldsymbol{\mathcal H}}^\mp(k_\text{pkt}) / x_\text{pkt}(k_\text{pkt}).
\end{align}
Let $k'$ denote the index of the first sample of an OFDM symbol associated to the tag sample $\hat{x}_\text{pkt}^\mp(k')$ of the oversampled tag data symbol. Then, the tag data sample associated to the OFDM symbols with timing instance $k'$ can be estimated by 
\begin{align}
\hat{x}_\text{pkt}^\mp(k') 
=
\left(\hat{\boldsymbol{\mathcal H}}_\text{pkt}^\mp\right)^\text{H}
\hat{\boldsymbol{\mathcal H}}^\mp(k').
\end{align}

Finally, maximum ratio combining may be used to combine the samples corresponding to the two bands and obtain the samples of an oversampled tag symbols. 
\begin{equation}
    \hat{x}_\text{pkt}(k') = \hat{x}_\text{pkt}^+(k') + \hat{x}_\text{pkt}^-(k')\exp(-\jmath\phi_\text{MRC}^\mp), \label{eq:tag_data_symbols}
\end{equation}
where $\phi_\text{MRC}^\mp$ is the combining phase obtained from multiple samples to align the signals corresponding to the lower and upper bands, respectively. 
For added processing gain, matched filtering can be employed to estimate the tag symbols from the sequential tag samples. 
Once the tag data symbols have been estimated, the estimated channels $\hat{\boldsymbol{\mathcal H}}^-$ and $\hat{\boldsymbol{\mathcal H}}^+$ can be de-rotated to perform the bistatic range estimation, described in the next section.

%% file: ranging.tex
\subsection{Range Information Retrieval}
Channel estimates $\hat{\boldsymbol{\mathcal H}}^\mp$ in \eqref{eq:ch_est+-spec} corresponding to the lower and upper band, respectively, need to be de-rotated by the estimated tag data symbols $\hat{x}_\text{pkt}(k_0)$ in \eqref{eq:tag_data_symbols}. Assuming correct tag data symbol detection, the frequency responses corresponding to the upper and lower band, respectively, may be written as $\hat{\boldsymbol{\mathcal H}}_{1,2}^\mp = \hat{\boldsymbol{\mathcal H}}^\mp \hat{x}_\text{pkt}(k_0)$, or using \eqref{eq:ch_est+-spec}
\begin{equation}
\hat{\boldsymbol{\mathcal H}}_{1,2}^\mp  = 
\mp\frac{\jmath}{\pi}\exp[\jmath(\phi_{\text{RX}}^\mp-\phi_{\text{TX}})] 
\boldsymbol{\mathcal D}_\zeta^0 \diag{\boldsymbol{\mathcal G}^\mp} 
(\boldsymbol{\mathcal H}_1 \odot \boldsymbol{\mathcal H}_2^\mp) 
+ \boldsymbol{\mathcal V}^\mp.  \label{eq:H_mp_est}
\end{equation}
The channel impulse responses corresponding to the central, lower and upper band are then obtained from the corresponding channel frequency responses in 
\eqref{eq:ch_est_0_spec}, \eqref{eq:ch_est+-spec}, respectively
by taking a larger \ac{IFFT} of size $N' > N$ (we assume that $N'$ is an even number) to achieve oversampling, hence increase the time granularity
\begin{eqnarray}
\hat{\bf h}_0 &=& {\mathbf F}_{N'}^\text{H}
[{\mathbf 0}_{Z}^\T, \:
(\hat{\boldsymbol{\mathcal H}}_0)^\T, \: 
{\mathbf 0}_{Z-1}^\T]^\T \label{eq:h_0_init} \\
\hat{\bf h}_{1,2}^\mp &=& {\mathbf F}_{N'}^\text{H}
[{\mathbf 0}_{Z}^\T, \:
(\hat{\boldsymbol{\mathcal H}}_{1,2}^\mp)^\T, \: 
{\mathbf 0}_{Z-1}^\T]^\T, \label{eq:h_1_2_init}
\end{eqnarray}
where $Z=(N'-N+1)/2$.
Note that due to limited illuminator waveform bandwidth, the estimated CIR taps will be widened or ``blurred" in time domain. The discrete channel taps are convolved with a Dirichlet kernel whose width is inversely proportional to the TX waveform bandwidth. Oversampling only increases ranging granularity and not resolution because the same blurred envelope is being oversampled.

\subsubsection{The central band channel}
Combining the frequency-domain estimated channel in \eqref{eq:ch_est_0_spec} and the true channel corresponding to the central band in \eqref{eq:H_0}, we may rewrite the ``blurred" impulse response of the central band due to limited bandwidth in \eqref{eq:h_0_init} as
\begin{equation} 
\hat{\bf h}_0 = {\mathbf \Phi}_0\boldsymbol{\mathcal\alpha}_0 + {\mathbf v}^0, \label{eq:h0_blurred}
\end{equation}
where ${\mathbf v}^0 = [v^0(0),\ldots,v^0(N'-1)]^\T$ is the $N'$-point \ac{IFFT} of the noise vector $\boldsymbol{\mathcal V}^0$ in \eqref{eq:ch_est_0_spec},
and
\begin{equation}
{\mathbf \Phi}_0 = a_0
{\mathbf F}_{N',N}
\boldsymbol{\mathcal D}_\zeta^0
\diag{\boldsymbol{\mathcal G}^0}
{\boldsymbol{\mathcal{F}}_0}  
{\boldsymbol{\mathcal{C}}_0} \label{eq:blurring_matrix}
\end{equation}
is the channel blurring matrix which relates the true channel impulse response taps $\boldsymbol{\mathcal\alpha}_0$ to the observed channel impulse response $\hat{\bf h}_0$ after oversampling. 
The complex-valued scalar $a_0$ in \eqref{eq:D_zeta_0} is $a_0=(1/2)\exp[\jmath(\phi_{\text{RX}}-\phi_{\text{TX}})]$.
Matrix ${\mathbf F}_{N',N}^\text{H}$ is the ${\mathbf F}_{N'}$ \ac{IFFT} matrix truncated to the middle $N$ columns, i.e., ${\mathbf F}_{N',N}^\text{H}={\mathbf F}_{N'}^\text{H}[:,-\frac{N-1}{2}:\frac{N-1}{2}]$.
Diagonal matrices $\boldsymbol{\mathcal D}_\zeta^0$ in \eqref{eq:D_zeta_0} and
$\diag{\boldsymbol{\mathcal G}^0}$ introduce linear phase rotations across subcarriers due to the OFDM timing delay $\zeta_0$, and due to the overall normalized group delay of the TX-RX circuitry 
$\tau_\text{g}$, respectively. 
Diagonal matrix ${\boldsymbol{\mathcal{C}}_0}$ given in \eqref{eq:rotation_tau_all_i} (for $i=0$)  introduces a similar linear phase rotation across subcarriers. Assuming that the TX-RX circuitry gain has been pre-calibrated, and its effect is only seen as a group delay, 
we may write the $(n',l)$ entries of ${\mathbf \Phi}_0$ as\par\noindent
\begin{align}  
\{{\mathbf \Phi}_0\}_{n',l} =& 
\exp\left(-\jmath 2 \pi \frac{F_\text{c}}{N\Delta F}\tau_{0}^l\right) \cdot \nonumber \\
&\cdot \frac{a_0}{\sqrt{N'}} \cdot
\frac{\sin\left[\pi N \left(\frac{n'}{N'}-\frac{\tau_{0}^l+\hat\zeta^0+\tau_\text{g}}{N}\right)\right]}{\sin\left[\pi\left(\frac{n'}{N'}-\frac{\tau_{0}^l+\hat\zeta^0+\tau_\text{g}}{N}\right)\right]}, \label{eq:Phi_0}
\end{align}
where $n'=0,\ldots,N'-1$ is the sample index at oversampled rate, and $l=0,\ldots,L_0-1$ is the propagation path index.
\begin{figure}
\centering
\includegraphics[width=.9\columnwidth]{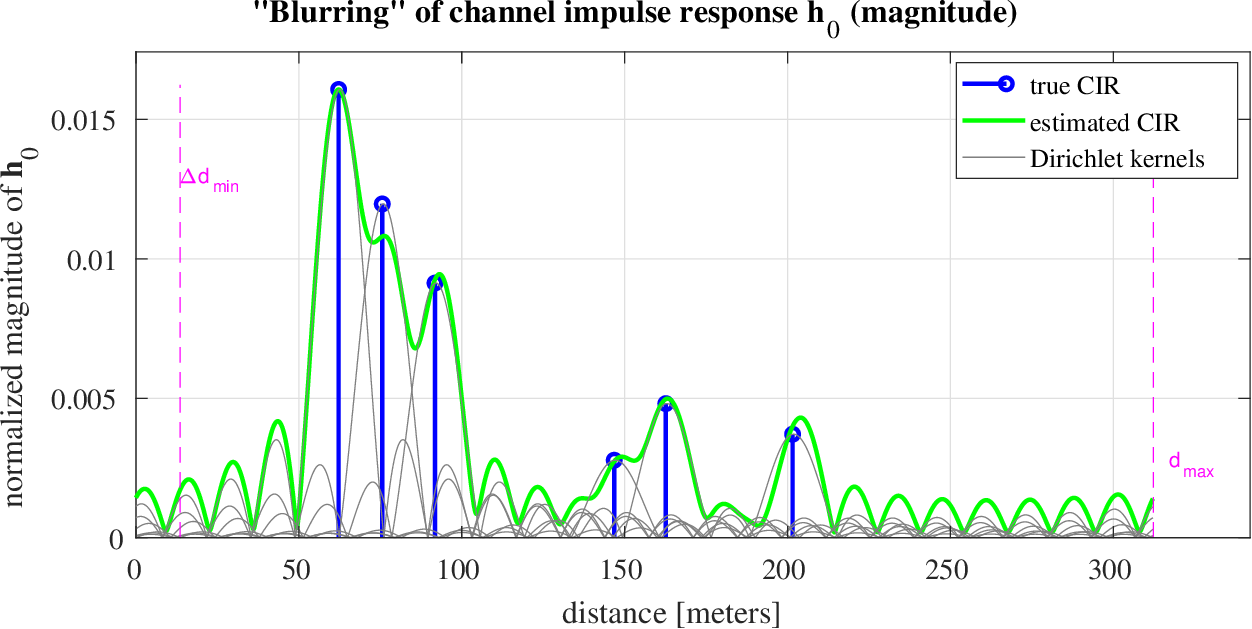}
\caption{The CIR ``blurring" due to limited bandwidth (shown by the thick green curve).
The channel taps (blue stem plot) are convolved with Dirichlet kernels (gray lines) whose widths are inversely proportional to the TX waveform bandwidth. Oversampling only increases ranging granularity, and not resolution because the same blurred envelope is being oversampled.
\label{fig:CIR_blurring} }
\end{figure}

Blurring matrix ${\mathbf \Phi}_0$ in \eqref{eq:Phi_0} plays the role of a "point spreading function" because it makes multipath components unresolvable in time domain if they are spaced at less than the inverse to the TX waveform bandwidth $B$.
It is important to note that oversampling via a larger \ac{IFFT} only increases the time granularity, and not the resolution that is the ability to separate closely-spaced multipath components, and which is dictated by the TX waveform bandwidth. Increasing granularity will sample the same CIR envelope which is a sum of Dirichlet kernels represented by the columns of ${\mathbf \Phi}_0$ in \eqref{eq:Phi_0}. The Dirichlet kernels cause the "blurring" of the channel taps, and their width is inversely proportional to TX waveform bandwidth, which means resolution can only be improved by increasing the illumination signal bandwidth. The CIR blurring phenomenon described by \eqref{eq:h0_blurred} for the direct TX-RX channel is shown in \figref{fig:CIR_blurring}, where the true CIR taps 
$\boldsymbol{\mathcal\alpha}_0$ are blurred by the corresponding Dirichlet kernels.

\subsubsection{The overall backscatter channels}
Following a similar derivation as for the central channel, we may write the overall backscatter channel impulse responses corresponding to the lower and upper band, respectively, as a concatenation of the TX-tag channel and the tag-RX channel, respectively. 
Replacing the expression of the true TX-RX channel $\boldsymbol{\mathcal H}_1$ in \eqref{eq:H_1} and the true tag-TX channel $\boldsymbol{\mathcal H}_2^\mp$ corresponding to the lower and upper bands in \eqref{eq:H_2-} and \eqref{eq:H_2+}, respectively into \eqref{eq:H_mp_est},
we may rewrite the CIR in \eqref{eq:h_1_2_init} corresponding to the backscattered lower and upper bands, respectively, after being ``blurred" due to the limited bandwidth  as
\begin{equation} 
\hat{\bf h}_{1,2}^\mp = 
({\mathbf \Phi}_1\boldsymbol{\mathcal\alpha}_1)
\ostar 
({\mathbf \Phi}_2^\mp\boldsymbol{\mathcal\alpha}_2^\mp)
+ {\mathbf v}^\mp, \label{eq:h_2_blurred}
\end{equation}
where ${\mathbf v}^\mp = [v^\mp(0),\ldots,v^\mp(N'-1)]^\T$ is the $N'$-point \ac{IFFT} of the noise vector $\boldsymbol{\mathcal V}^\mp$ in \eqref{eq:ch_est+-spec} and
\begin{align}
{\mathbf \Phi}_1 &= 
{\mathbf F}_{N',N}
{\boldsymbol{\mathcal{F}}_1}  
{\boldsymbol{\mathcal{C}}_1},
\label{eq:blurring_matrix_1} \\
{\mathbf \Phi}_2^\mp &= a_2^\mp
{\mathbf F}_{N',N}
\boldsymbol{\mathcal D}_\zeta^0
\diag{\boldsymbol{\mathcal G}^\mp}
{\boldsymbol{\mathcal{F}}_2}  
{\boldsymbol{\mathcal{C}}_2}
{\boldsymbol{\mathcal{D}}_2}^\mp
\label{eq:blurring_matrix_mp} 
\end{align}
are the channel ``blurring" matrices which relates the true channel impulse response taps to the observed channel impulse response after oversampling for the TX-tag channel, and lower and upper band of the tag-RX channel, respectively. 
Matrices ${\boldsymbol{\mathcal{D}}_2}^-={\boldsymbol{\mathcal{D}}_2}$ and ${\boldsymbol{\mathcal{D}}_2}^+={\boldsymbol{\mathcal{D}}_2}^\text{H}$, and ${\boldsymbol{\mathcal{D}}_2}$ is given in \eqref{eq:shift_tau_all_i}. 
The complex-valued scalar $a_2^\mp$ in \eqref{eq:blurring_matrix_1}--\eqref{eq:blurring_matrix_mp} is $a_2^\mp=\mp(\jmath/\pi)\exp[\jmath(\phi_{\text{RX}}^\mp-\phi_{\text{TX}})]$.
Diagonal matrix $\diag{\boldsymbol{\mathcal G}^\mp}$ introduce linear phase rotations across subcarriers due to the overall normalized group delay of the TX-tag-RX circuitry $\tau_\text{g}^\mp$ in the lower and upper bands, respectively. 
Diagonal matrices ${\boldsymbol{\mathcal{C}}_i}, i\in \{1,2\}$ given in \eqref{eq:rotation_tau_all_i} introduce similar linear phase rotations across subcarriers. Assuming that the TX-tag-RX circuitry gain has been pre-calibrated, and its effect is only seen as a group delay $\tau_\text{g}^\mp$ in the lower and upper bands, 
we can write the $(n',l)$ entries of ${\mathbf \Phi}_1$ and ${\mathbf \Phi}_2^\mp$ as
\begin{align}  
 &\{{\mathbf \Phi}_1\}_{n',l_1} = 
\exp\left(-\jmath 2 \pi \frac{F_\text{c}}{N\Delta F}\tau_{1}^{l_1}\right)\frac{\sin\left[\pi N \left(\frac{n'}{N'}-\frac{\tau_{1}^{l_1}}{N}\right)\right]}{\sin\left[\pi\left(\frac{n'}{N'}-\frac{\tau_{1}^{l_1}}{N}\right)\right]}, \label{eq:Phi_1} \\
 &\{{\mathbf \Phi}_2^\mp\}_{n',l_2} = 
\exp\left(-\jmath 2 \pi \frac{F_\text{c}\mp F_\text{shift}}{N\Delta F}\tau_{2}^{l_2}\right) \cdot  \nonumber \\
&\hspace{15mm} \cdot \frac{a_2^\mp}{\sqrt{N'}} \cdot
\frac{\sin\left[\pi N \left(\frac{n'}{N'}-\frac{\tau_{2}^{l_2}+\hat\zeta^0+\tau_\text{g}^\mp}{N}\right)\right]}{\sin\left[\pi\left(\frac{n'}{N'}-\frac{\tau_{2}^{l_2}+\hat\zeta^0+\tau_\text{g}^\mp}{N}\right)\right]}, \label{eq:Phi_2_mp}
\end{align}
where $n'=0,\ldots,N'-1$ is the sample index at oversampled rate, and $l_i=0,\ldots,L_i-1, \:
i \in \{1,2\}$ are the propagation path indices corresponding to TX-tag and tag-RX channels, respectively.
We validate the analytic \ac{CIR} expressions using the wireless emulator in \secref{sec:results}.

\subsection{Bi-static Range Estimation}
The objective of bistatic ranging is to estimate the cumulative distance between TX-tag-RX based on backscattered signal time-of-flight (ToF). Ensuring a tight time/clock synchronization between multiple TX-RX pairs is very costly and presents scalability issues in wide area deployments. A practical, scalable, and cost-effective approach, instead, is to utilize the direct TX-RX channel as a time reference for each TX-RX pair. 
Therefore, the bistatic range estimation relies on two timing measurements: 
1) the estimated \ac{ToF} on the compound TX-Tag-RX backscatter channel $\tau_1^0+\tau_2^0 + \zeta_0 + \tau_\text{g}^\mp$ and 
2) the estimated ToF on the direct TX-RX channel $\tau_0^0 + \zeta_0 + \tau_\text{g}^0$.


By subtracting the timing measurements, we obtain an estimate of the \textit{range-difference} $\Delta \hat{d}$
\begin{equation} \label{eq:range_diff}
\Delta \hat{d} = [(\tau_1^0+\tau_2^0)-\tau_0^0]c/B +d_\text{calib}^\mp,
\end{equation}
where $d_\text{calib}^\mp = (\tau_\text{g}^\mp-\tau_\text{g}^0)c/B$ is the calibration distance due to the difference in group delays in the corresponding bands due to transceiver circuitry which can be calibrated.

Finally, the bistatic range estimate $\hat{d}_{1,2}^\mp$ corresponding to the compound TX-tag-RX channel can be obtained by adding the known $d_0$ and subtracting the calibration distance $d_\text{calib}^\mp$ from $\Delta \hat{d}$ as follows:
\begin{equation}
\hat{d}_{1,2}^\mp =  \left[d_0 + \Delta\hat{d} - d_\text{calib}^\mp\right] \oplus d_\text{max}, \label{eq:est_bistatic_range}
\end{equation}
where $d_\text{max} = c/\Delta F$ is the maximum unambiguous range corresponding to the OFDM illumination waveform. 

Since the TX-RX channel is used as a reference (assuming the known true TX-RX distance $d_0$), at least two transceiver chains are required, one for the TX-RX channel (as a reference), and a second one for one of the backscatter sub-bands.
Having a third transceiver chain to cover both the lower and the upper backscatter sub-bands can provide frequency diversity, hence resilience to the frequency selectivity and possibility to combine the range estimates from both sub-bands. These can be combined based on the reliability metrics obtained from the raw signals on each sub-band, such as \ac{SINR} or \ac{BER}
\begin{equation}
\hat{d}_{1,2} =  \frac{1}{w^- + w^+} (w^- \hat{d}_{1,2}^- + w^+ \hat{d}_{1,2}^+),  \label{eq:est_bistatic_range_combined}
\end{equation} 
where $w^-,w^+$ are the reliability metrics. 
\subsection{IR FIRST: Peak search-based bi-static range estimation}
In this section, we describe a ranging method, called \textit{IR First}, which uses the first peak of CIRs for bi-static ranging. 
Let the index of the first peak of the oversampled IR of TX-RX channel above a predefined magnitude threshold $a_\text{min}$ is
\begin{align}
i_0 = \min \Big\{ n' \: \Big| \: 
&|\hat{h}_0(n'-1)| < |\hat{h}_0(n')| \geq |\hat{h}_0(n'+1)|, \nonumber \\
&|\hat{h}_0(n')| \geq a_\text{min} \Big\}, \label{eq:ind_first_peak_0}
\end{align}
where $\hat{h}_0(n'), \: n'=1,\ldots,N'$ are the elements of the oversampled CIR vector 
$\hat{\bf h}_0$ in \eqref{eq:h0_blurred}. Similarly, the first significant peaks of the lower and upper band \ac{CIR}s are
\begin{align}
i_{1,2}^\mp = \min \Big\{ n' \: \Big| \: 
&|\hat{h}_{1,2}^\mp(n'-1)| < |\hat{h}_{1,2}^\mp(n')| \geq |\hat{h}_{1,2}^\mp(n'+1)|, \nonumber \\
&|\hat{h}_{1,2}^\mp(n')| \geq a_\text{min}, n' > i_0 \Big\}, \label{eq:ind_first_peak_mp}
\end{align}
where $\hat{h}_{1,2}^\mp(n'), \: n'=1,\ldots,N'$ are the elements of the CIR vectors 
$\hat{\bf h}_{1,2}^\mp$ in \eqref{eq:h_2_blurred}. Then, the range-difference estimate $\Delta \hat{d}$ and the bi-static range estimate $\hat{d}_{1,2}^\mp$ using the IR First method are given by 
\begin{equation}
    \Delta \hat{d}  =  \frac{(i_{1,2}^\mp - i_0)c}{N'\Delta F}. \label{eq:est_bistatic_range_diff_IR_FIRST}
\end{equation}
\\
The estimated compound TX-Tag-RX bistatic range is
\begin{equation}
    \hat{d}_{1,2}^\mp  =  \big[d_0 + \frac{(i_{1,2}^\mp - i_0)c}{N'\Delta F} - d_\text{calib}^\mp\big] \oplus d_\text{max}. \label{eq:est_bistatic_range_IR_FIRST}
\end{equation}
%
%
Note that, with IR First, the granularity of $\hat{d}_{1,2}^\mp$ is limited by the granularity of $\Delta \hat{d}$, which, in turn, is driven by the IFFT size $N'$. Since $i_0, i_{1,2}^\mp$ are integers, the finest granularity in the estimate $\hat{d}_{1,2}^\mp $ is $2c/(N'\Delta F)$.
\subsection{Cramér–Rao Lower Bound}
Bistatic CRLB has been addressed in the multicarrier radar context in ~\cite{bica2017delay,bica2018radar}. In this paper, we address the CRLB of frequency-shifted bistatic backscatter range estimation. 
We now derive the CRLB for the estimate $\hat{d}^{+}_{1,2}$ given the estimated TX-RX channel $\hat{\boldsymbol{\mathcal H}}_0$ in \eqref{eq:ch_est_0_spec} and the estimated upper-band backscatter channel $\hat{\boldsymbol{\mathcal H}}_{1,2}^{+}$ in \eqref{eq:H_mp_est} as the observations. 
We derive the bound in two steps: First, we determine the CRLB of the TX-RX \ac{LoS} distance ${d}_0$ and the TX-Tag-RX distance ${d}_{1,2}$. Second, we describe the final CRLB as a sum of the two bounds, assuming the independence between the channel taps of TX-RX and TX-Tag-RX channels.

The following derivation is provided only for the upper band, but can be readily adapted to the lower band. 

\subsubsection{CRLB for the TX-RX range estimate}
For the direct TX-RX channel range estimation, we consider the \ac{LoS} path length $d_0$ as parameter of interest, whereas the path lengths of the other multipath components of the channel as the nuisance parameters. 
We denote the vector these \ac{NLoS} path lengths as
\begin{equation}
\mathbf{d}^0_\text{NLoS} = [d_0^1,\ldots,d_0^{L_0-1}]^\T.
\end{equation}

We rewrite  $\hat{\boldsymbol{\mathcal H}}_0$ in \eqref{eq:ch_est_0_spec} as 

\begin{equation}
\hat{\boldsymbol{\mathcal H}}_0 = \boldsymbol{\mu}_0+ \boldsymbol{\mathcal{V}}^{0}, 
\quad \boldsymbol{\mathcal{V}}^{0} \sim \mathcal{CN}(0, \mathbf{Q}_0),
\end{equation}
where 
\begin{align}\label{eq:mu-0}
\boldsymbol{\mu}_0 = + \frac{1}{2} 
\exp[{\jmath (\phi_\text{RX}-\phi_\text{TX})}]
\boldsymbol{\mathcal{D}}_\zeta^0 \diag{\boldsymbol{\mathcal{G}}^0} 
\boldsymbol{\mathcal{H}}_0.
\end{align}
Then, the likelihood of $(d_0, \mathbf{d}^0_\text{NLoS})$ is given by
\begin{align}
&\mathcal{L}(d_0,\mathbf{d}^0_\text{NLoS}) 
=p\big(\hat{\boldsymbol{\mathcal H}}_0 \,\big|\, d_0,\mathbf{d}^0_\text{NLoS} \big) 
\\
&=\frac{1}{\pi^N \det(\mathbf{Q}_0)} 
\exp \Big[
- \big(\hat{\boldsymbol{\mathcal H}}_0 - \boldsymbol{\mu}_0\big)^\text{H}
\mathbf{Q}_0^{-1}\big(\hat{\boldsymbol{\mathcal H}}_0 - \boldsymbol{\mu}_0 \big)
\Big].\nonumber
\end{align}
The associated Fisher information can be defined as ~\cite{kay1993statistical}
\begin{equation} \label{eq:FI_0}
\mathcal{I}(d_0)=-\mathbb{E}_{{\mathbf d}^0_\text{NLoS}}\!\left
\{\mathbb{E}_{{{\boldsymbol{\mathcal{V}}}^0}}\left\{\frac{\partial^2 \ln \mathcal{L}(d_0,\mathbf{d}^0_\text{NLoS})}{\partial d_0^2}\Big | \mathbf{d}^0_\text{NLoS} \right\}\right \}.
\end{equation}
Using $\mathbb{E}_{{{\boldsymbol{\mathcal{V}}}^0}}\left\{\hat{\boldsymbol{\mathcal{H}}}_0 - \boldsymbol{\mu}_0\right\}=\boldsymbol{0}$ and $\tau_0=\frac{d_0B}{c}$, \eqref{eq:FI_0} reduces to
\begin{align} \label{eq:FI_0_again}
\mathcal{I}(d_0)
&= \frac{2B^2}{c^2} \mathbb{E}_{{\mathbf d}^0_\text{NLoS}}\!
\left\{  
\frac{\partial\boldsymbol{\mu}_0}{\partial\tau_0}^{\H}
\mathbf{Q}_0^{-1}
\frac{\partial \boldsymbol{\mu}_0}{\partial \tau_0 } 
\right\}.
\end{align}
Finally, assuming $\mathbf{Q}_0 = \sigma_0^2\mathbf{I}$,
\begin{align}
\text{CRLB}(d_0)
&=\mathcal{I}(d_0)^{-1} = \frac{c^2 \sigma_0^2}{2B^2} \,
\mathbb{E}_{{\mathbf d}^0_\text{NLoS}}\!\Big\{ \Big\| 
\frac{\partial\boldsymbol{\mathcal{H}}_0}{\partial \tau_0^0}
\Big\|_2^2\Big\}^{-1}\nonumber \\
&= \frac{c^2 \sigma_0^2}{2B^2 {\pi^2 |\alpha_0^0|^2}} 
\Big(\frac{F_\text{c}^2}{B^2}+\frac{N^2-1}{12N}\Big)^{-1}. \label{eq:CRLB_final_0}
\end{align}
\vspace{-1ex}
\subsubsection{CRLB for the TX-Tag-RX range estimate}
The TX-Tag-RX channel is a convolution of the TX-Tag channel and the Tag-RX channel. Therefore, we consider it to have $L_1L_2$ distinct channel paths. Among these paths, we consider the direct path length $d_{1,2}=d_1+d_2$ as a parameter of interest, and the path lengths of the rest channel paths as the nuisance parameters. We denote these parameters as 
\begin{align}
\mathbf{d}^{1,2}_\text{NLoS}&=[d_1^{0}+d_2^{1},\ldots,d_1^{0}+d_2^{L_2-1},d_1^{1}+d_2^{0},\ldots,d_1^{1}+
d_2^{L_2-1}, \nonumber \\
&,\ldots, d_1^{L_1-1}+d_2^{0},\ldots,d_1^{L_1-1}+d_2^{L_2-1}]^\T.
\end{align} 
We rewrite  
$\hat{\boldsymbol{\mathcal H}}_{1,2}^{+}$ in \eqref{eq:H_mp_est} as 
\begin{equation}
\hat{\boldsymbol{\mathcal H}}_{1,2}^{+} = \boldsymbol{\mu}^{+}+ \boldsymbol{\mathcal{V}}^{+}, 
\quad \boldsymbol{\mathcal{V}}^{+} \sim \mathcal{CN}(0, \mathbf{Q}^+_{1,2}),
\end{equation}
where 
\begin{align}\label{eq:mu-plus}
\boldsymbol{\mu}^{+} = \frac{\jmath}{\pi} 
\exp[{\jmath (\phi_\text{RX}^+-\phi_\text{TX})}]
\boldsymbol{\mathcal{D}}_\zeta^0 \diag{\boldsymbol{\mathcal{G}}^{+}}
(\boldsymbol{\mathcal{H}}_1 \odot \boldsymbol{\mathcal{H}}_2^{+}).
\end{align}
Therefore, the likelihood of $(d_{1,2},\mathbf{d}^{1,2}_\text{NLoS})$ is given by
\begin{align}
&\mathcal{L}(d_{1,2},\mathbf{d}^{1,2}_\text{NLoS}) 
=p\big(\hat{\boldsymbol{\mathcal H}}_{1,2}^{+} \,\big|\, d_{1,2},\mathbf{d}^{1,2}_\text{NLoS} \big) 
\nonumber\\
&=\frac{1}{\pi^N \det(\mathbf{Q}^+_{1,2})} 
\exp [
- (\hat{\boldsymbol{\mathcal H}}_{1,2}^{+} - \boldsymbol{\mu}^{+}\big)^\text{H}
(\mathbf{Q}^+_{1,2})^{-1}(\hat{\boldsymbol{\mathcal H}}_{1,2}^{+} - \boldsymbol{\mu}^{+} \big)
].
\end{align}
The associated Fisher information can be defined as~\cite{kay1993statistical}
\begin{equation} \label{eq:FI_12}
\mathcal{I}(d_{1,2})=-\mathbb{E}_{\mathbf{d}^{1,2}_\text{NLoS}}\!\left
\{\mathbb{E}_{{\boldsymbol{\mathcal{V}}}^{+}}\!\left\{\frac{\partial^2 \ln \mathcal{L}(d_{1,2},\mathbf{d}^{1,2}_\text{NLoS})}{\partial d_{1,2}^2}\Big | \mathbf{d}^{1,2}_\text{NLoS} \right\}\right \}.
\end{equation}
Again, $\mathbb{E}_{{\boldsymbol{\mathcal{V}}}^{+}}\left\{\hat{\boldsymbol{\mathcal{H}}}_{1,2}^{+} - \boldsymbol{\mu}^{+}\right\}=\boldsymbol{0}$ and $\tau_{(\cdot)}=\frac{d_{(.)}B}{c}$ gives
\begin{align} \label{eq:FI_12_again}
\mathcal{I}(d_{1,2})
&= \frac{2B^2}{c^2} \mathbb{E}_{{\mathbf d}^{1,2}_\text{NLoS}}
\left\{  
\frac{\partial\boldsymbol{\mu}^{+}}{\partial\tau_\text{1,2}}^{\H}
(\mathbf{Q}_{1,2}^+)^{-1}
\frac{\partial \boldsymbol{\mu}^{+}}{\partial \tau_\text{1,2} } 
\right\},
\end{align}
where $\tau^0_\text{1,2}=\tau_1^{0}+\tau_2^{0}$. Finally, assuming $\mathbf{Q}_{1,2}^+ = \sigma_{1,2}^2\mathbf{I}$,
\begin{align}
\text{CRLB}(d_\text{1,2})
&=\mathcal{I}(d_\text{1,2})^{-1}  \\
&= \frac{c^2 \sigma_{1,2}^2 \pi^2}{2B^2} \,
\mathbb{E}_{{\mathbf d}^{1,2}_\text{NLoS}}\!\Big\{ \Big\| 
\frac{\tau_1^0}{\tau^0_\text{1,2}} 
\left(
\frac{\partial\boldsymbol{\mathcal{H}}_1}{\partial \tau_1^0}\odot 
\boldsymbol{\mathcal{H}}_2^{+}
\right) + \nonumber\\
&+\frac{\tau_2^0}{\tau^0_\text{1,2}} 
\left(
\boldsymbol{\mathcal{H}}_1 \odot \frac{\partial \boldsymbol{\mathcal{H}}_2^{+}}{\partial \tau_2^0} \right)
\Big\|_2^2\Big\}^{-1}.\label{eq:CRLB_final_12}
\end{align}
\subsubsection{CRLB on the final range estimate}
We assume the direct TX-RX channel and the compound backscatter TX-Tag-RX channel path lengths to be independent. Therefore, based on \eqref{eq:range_diff}, the CRLB of the upper band bistatic range $d^+_{1,2}$ is the sum of CRLB of $d_0$ in \eqref{eq:CRLB_final_0} and the CRLB of $d_{1,2}$ in \eqref{eq:CRLB_final_12}.
\begin{equation}
    \text{CRLB}(d^+_{1,2}) = \text{CRLB}({d}_0) + \text{CRLB}({d}_\text{1,2}).
\end{equation}
Note that the derived CRLB results require prior knowledge of the channel NLoS components. Therefore, in our experiments, we evaluate a stochastic CRLB by averaging the Fisher information over uniformly random realizations of the scatterers; positions, producing the NLoS components $\mathbf{d}^+_\text{NLoS}$ and $\mathbf{d}^0_\text{NLoS}$.

%% file: results.tex
In this section, we first validate the expressions of the analytical CIRs derived in \eqref{eq:h0_blurred} and \eqref{eq:h_2_blurred} by comparing them with the experimentally estimated CIRs from a full-scale emulation setup. We then use the emulation setup to present the limits of the bi-static backscatter system in terms of the ranging and communication performance as a function of the multipath and the SNR. 
\subsection{Channel Generation}
\subsubsection{Geometry}
The illuminator and the receiver are located at $(-8,0) $m and $(8,0)$ m, respectively. 
For each scenario, a position of the tag is randomly selected from the region $[-9,9] \times [-8,8] m^2$. Subsequently, the specified number of scatterers are also chosen uniformly randomly with the mean at the centroid of TX, Tag and RX positions. 

\subsubsection{Channel Parameters}
Based on the geometry, the path lengths and path delays $(d_i^l, t_i^l), l\in\{0,...,L_i-1\}, i\in\{0,1,2\}$ are determined. The channel coefficients $\alpha_i^l,l\in\{0,...,L_i-1\}, i\in\{0,1,2\}$ are generated with the magnitude inversely proportional to the path lengths, and uniformly random phases.

Unless otherwise mentioned, we maintain a geometrically consistent channel. Therefore, we assume that all three channels of the bi-static backscatter system have same number of multipath components (albeit with varying relative path loss), resulting from the number of scatterers in the environment. Therefore, $L_0 = L_1 = L_2 \geq 1.$

\subsection{Emulating a Bi-static Backscatter System}
\begin{figure*}
    \centering
    \includegraphics[width=.85\linewidth]{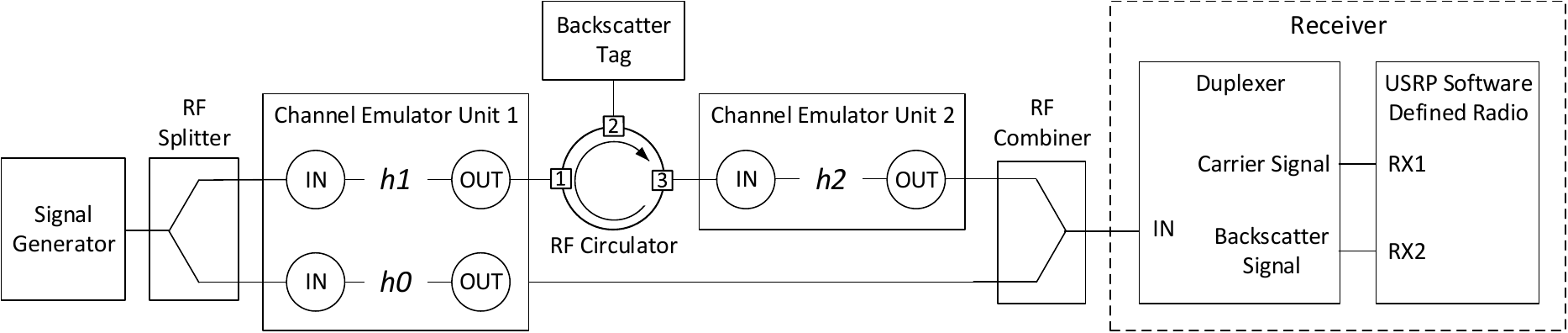}
    \caption{Emulating a bi-static backscatter system with Keysight MXG as an illuminator, a custom-designed backscatter tag, a USRP Software Defined Radio receiver and two Spirent SR5500 RF channel emulators.\vspace{-1in}}%
    \label{fig:EmulationSetup}
\end{figure*}

The emulation setup is illustrated in \figref{fig:EmulationSetup}. The primary components include two Spirent SR5500 channel emulators, which provide control over the delay, phase, and relative path loss for up to 12 paths per channel. Using two channel emulators, we can independently emulate all three channels -- TX-RX, TX-Tag and Tag-RX channels -- of the bi-static backscatter system.

\subsubsection{Operation of the emulation setup}
The illuminator signal is generated using a Keysight MXG signal generator, which produces repeated OFDM symbols with $N=23$ subcarriers, spaced at $\Delta F=960$ kHz, centered at $F_\text{c}=897.5$ MHz.

The illumination signal is divided into two paths using a splitter: 
(1) One path passes through an emulated TX-RX channel and is received by the receiver, serving as the reference signal.
(2) The other path passes through an emulated TX-Tag channel, with the resultant signal directed to the backscatter tag using a circulator.  

The backscatter tag reflects the input signal, shifting it by $F_\text{shift} = 45$ MHz. The tag operates in two modes: (1) CW mode, where the tag continuously reflects the signal without any modulation, and (2) Burst mode, where the tag modulates a tag packet once every 50 ms cycle~\cite{2025_PatZhaKim}. 
The signal reflected from the tag, now centered at $F_c+F_\text{shift}$, passes through the channel emulator which emulates the Tag-RX channel.

The signals from the TX-RX channel ($h_0$) and the Tag-RX channel ($h_2$) are combined to form the input signal to the receiver. The receiver consists of two stages. In the first stage, the input signal is split into two bands, centered at $F_c$ and $F_c+F_\text{shift}$, using a duplexer. These outputs are fed into two channels of USRP X410, centered at the respective frequencies.

\subsubsection{Receiver Processing}
The USRP X410 processes two signals separately. It samples the signals at 62.5 \ac{MSps}, which are then digitally resampled to $F_\text{samp}=61.44$ MSps. The resampled signals undergo further processing as detailed in \secref{sec:reader} to finally provide experimentally estimated CIRs $\hat{\mathbf{h}}_0$ and $\hat{\mathbf{h}}_{1,2}^{\mp}$.

This emulation setup allows for control of all three channels in a bi-static backscatter system and also incorporates real-world RF impairments expected at the receiver. Prior to experiments, the system is calibrated by setting the tag in CW mode, configuring all emulated channels as \ac{LoS} with zero delays, and synchronizing all units. The \ac{CFR} of both TX-RX and TX-Tag-RX channels are estimated at the receiver. This calibrated frequency response is used to compensate for all subsequent measurements.
\subsection{Validating Analytic Model}
We validate the analytic model of \ac{CIR}s derived in \eqref{eq:h0_blurred} and \eqref{eq:h_2_blurred} by comparing them with the experimentally estimated \ac{CIR}s. 
To facilitate comparison of the analytical \ac{CIR}s of the TX-RX and the TX-Tag with the corresponding estimated CIRs from the emulator, we synchronize the signal generator, the channel emulators and the receiver. 
We then normalize and time align the analytical CIRs and experimentally estimated CIRs. 
Specifically, both CIRs are normalized by the maximum of their magnitude in order to remove the CIR scaling due to analog and digital scaling:
\begin{equation} \label{eq:norm_CIRs}
    \hathat{h}_{\{\cdot\}}(n') = \frac{\hat{h}_{\{\cdot\}}(n')}{\max_{n''} 
    |\hat{h}_{\{\cdot\}}(n'')|}, \quad n',n''=0,\ldots,N'-1,
\end{equation}
where subscript $\{\cdot\}$ is a placeholder to denote the direct TX-RX channel $\hathat{h}_0(n')$, and the combined TX-Tag-RX channels $\hathat{h}_{1,2}^\mp(n')$ corresponding to the lower and upper bands, respectively. 
Then, the normalized CIRs are time-aligned solely based on the position of their maxima (no least squares matching is performed). 
The CIR measured in the experimental setup $\hathat{h}'_{\{\cdot\}}(n')$ 
is circularly shifted to compensate for the overall group delay $m$ in the hardware setup rounded up to sampling granularity, such that its maximum index matches the maximum index of 
analytically derived CIR $\hathat{h}_{\{\cdot\}}(n')$, i.e.,
\begin{equation}
    \hathat{h}''_{\{\cdot\}}(n') = \hathat{h}'_{\{\cdot\}}[(n'+m)\oplus N'], \quad n'=0,\ldots,N'-1. 
\end{equation}

After scaling and alignment, we may define the point-wise \ac{NMAE} between the analytic CIR vector $\hathat{\mathbf{h}}_{\{\cdot\}}$ and the estimated CIR vector $\hathat{\mathbf{h}}''_{\{\cdot\}}$ from the emulated measurement as follows: 
\begin{equation}
    \text{NMAE}(\hathat{\mathbf{h}}_{\{\cdot\}}, \hathat{\mathbf{h}}''_{\{\cdot\}}) 
    = \frac{\mathbf{1}_{N'}^\T \left\vert\hathat{\mathbf{h}}_{\{\cdot\}} - \hathat{\mathbf{h}}''_{\{\cdot\}}\right\vert}{\mathbf{1}_{N'}^\T \left\vert\hathat{\mathbf{h}}''_{\{\cdot\}}\right\vert}.
\end{equation}

\begin{figure}[t]
\centering
\subfloat[TX-RX channel: NMAE = 0.0211 \label{fig:CIR_scene1_0}]{\includegraphics[width=.85\linewidth]{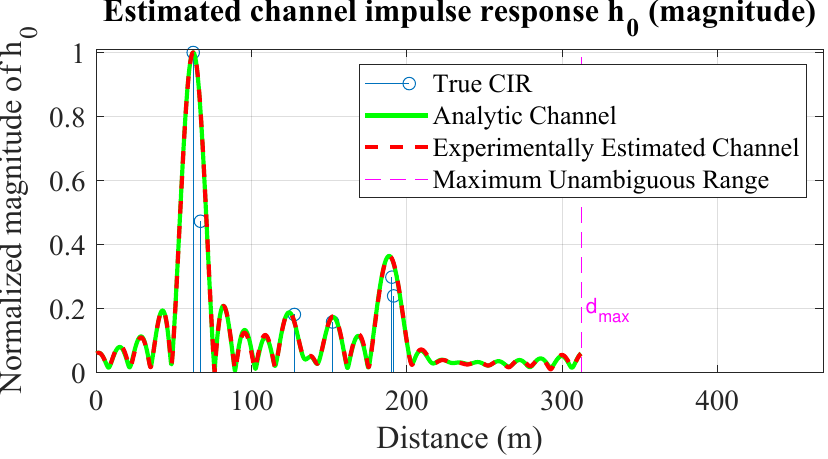}}\\
\subfloat[TX-Tag-RX channel: NMAE = 0.0610 \label{fig:CIR_scene1_1_2}]{\includegraphics[width=.85\linewidth]{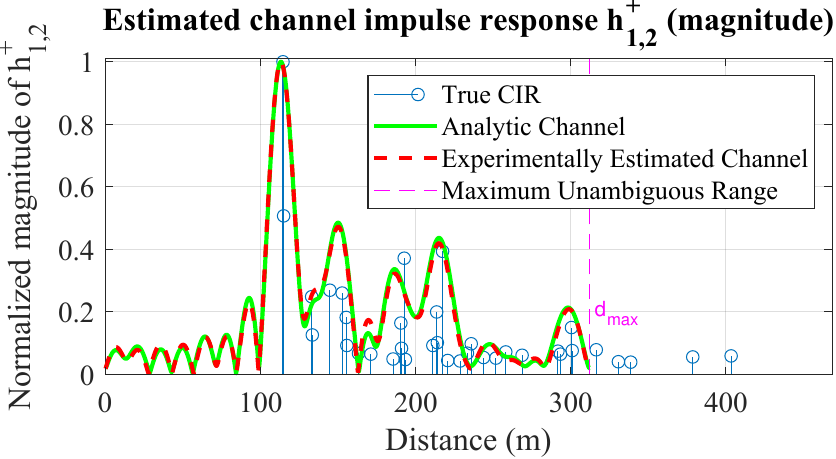}}
\caption{Comparison of analytical and emulated CIRs: The apparent similarity of analytical and experimentally estimated CIRs show the validity of the analytic expressions derived in \eqref{eq:h0_blurred} and \eqref{eq:h_2_blurred}. Average NMAE across 200 randomly generated channels are given in \tabref{tbl:NMAE_analytic_emulated}.}
\label{fig:CIRComparison}
\end{figure}

\figref{fig:CIRComparison} shows the good match between the analytic CIRs and the emulated CIRs with NMAE of 0.0211 and 0.0610, respectively. Note that despite the fact that $h_{1,2}^+(t)$ in \figref{fig:CIRComparison} has 36 taps, the analytical CIR matches almost perfectly the measured CIR. 
We further test the similarity by considering 200 random channel realizations with $L_{i}=\{1,3,5,7,9\}$ paths in all three channels. 
Analytical and experimentally estimated CIRs corresponding to each channel realization are compared. The average NMAEs are presented in \tabref{tbl:NMAE_analytic_emulated}. The average NMAE is less than 0.035 for the TX-RX channel $\hat{\mathbf{h}}_0$, and less than 0.12 for the TX-Tag-RX channel $\hat{\mathbf{h}}^+_{1,2}$, irrespective of the number of paths, which shows the validity of the analytical model.    
\begin{table}[t]
\centering
\caption{Comparison of Analytical and Experimentally Estimated CIRs}\label{tbl:NMAE_analytic_emulated}
\begin{tabular}{c|c|c} 
\hline
\begin{tabular}[c]{@{}c@{}}\textbf{Number of }\\\textbf{Scatterers}\end{tabular} & \begin{tabular}[c]{@{}c@{}}\textbf{Normalized}\\\textbf{MAE of $\hat{\mathbf{h}}_0$}\end{tabular} & \begin{tabular}[c]{@{}c@{}}\textbf{Normalized}\\\textbf{MAE of $\hat{\mathbf{h}}_{1,2}^+$}\end{tabular} \\ 
\hline
0 & 0.0234 & 0.1039 \\ \hline
2 & 0.0227 & 0.0993 \\ \hline
4 & 0.0288 & 0.1191 \\ \hline
6 & 0.0312 & 0.1175 \\ \hline
8 & 0.0275 & 0.1068 \\ \hline
\end{tabular}
\end{table}
\subsection{Resolvability of Multipath Components}
\begin{figure}[t]
\centering
\subfloat[TX-RX channel \label{fig:DelayVsAmplitude_h0}]{\includegraphics[width=.85\linewidth]{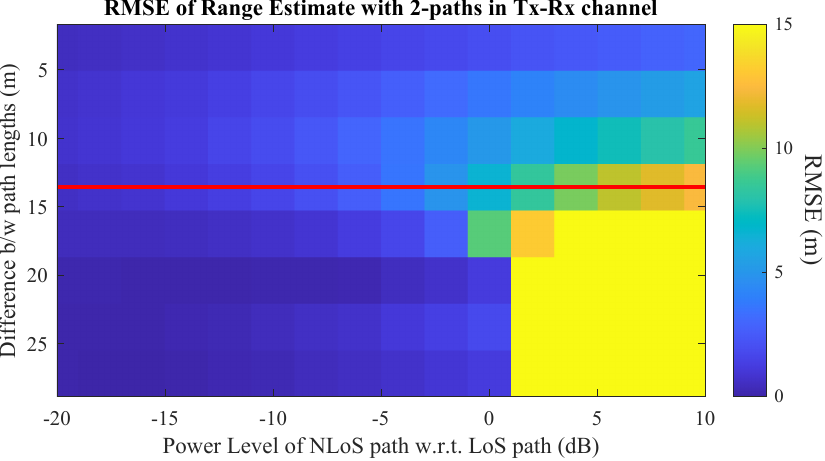}}\\
\subfloat[Tag-Rx channel \label{fig:DelayVsAmplitude_h2}]{\includegraphics[width=.85\linewidth]{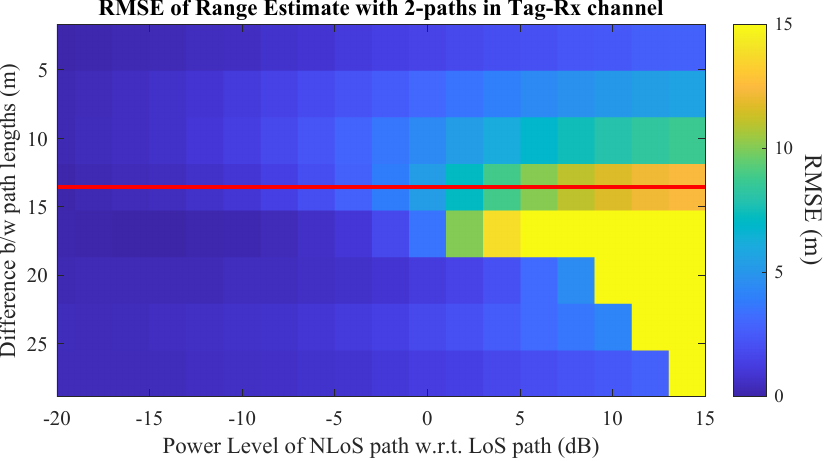}}
\caption{\ac{RMSE} of the range estimate as function of the relative path length and the relative power of the  path. Brighter color represents higher error in the range estimate. The red line indicates the theoretical resolvability limit of $c/B=13.58$m. As the difference between path lengths moves away from the resolvability limit for small enough power of NLoS component, identifying the LoS path gets easier, hence, the range estimation error reduces. As the power of NLoS path increases, differentiating the LoS path from the noise gets challenging, resulting in higher range estimation error.}
\label{fig:DelayVsAmplitude}
\end{figure}

In this section, we use the emulation setup to showcase the fundamental limit on the multipath resolvability and its impact on the range estimation.
To understand the multipath resolvability, we generate an arbitrary channels with two paths: a LoS path and a NLoS path. We fix the LoS path length and power, and then vary the NLoS path length and the power. For each value of the NLoS path length and the power of the NLoS path, we estimate the bi-static range $\hat{d}_{1,2}$ using the IR First method with $N'=4096$ as described in \secref{sec:ranging}. 
We limit our analysis to the high SNR setting to avoid the effect of noise on the ranging error and isolate the cause of the observed ranging error exclusively to the NLoS paths. 

\figref{fig:DelayVsAmplitude} shows the range estimation \ac{RMSE} as a function of the relative path length and the power of the NLoS path with respect to the LoS path. Specifically, in \figref{fig:DelayVsAmplitude_h0}, we consider TX-RX channel with two paths, and other channels have only LoS path, i.e., $L_0=2,L_1,L_2=1$, implying the ranging error is primarily due to the NLoS component in TX-RX channel. Similarly, in \figref{fig:DelayVsAmplitude_h2}, we consider only Tag-RX channel to be affected by NLoS channel, i.e., $L_0,L_1=1,L_2=2$. 
\figref{fig:DelayVsAmplitude} shows that when the difference between LoS and NLoS path lengths is close to the theoretical resolvability limit $c/B$ (indicated by red lines in \figref{fig:DelayVsAmplitude}), even a low-power NLoS path contributes to the error in the range estimate. As the difference between the LoS and NLoS path lengths move away from the theoretical resolvability limit, the LoS path can be resolved and hence, the range estimation error reduce. Furthermore, differentiating the LoS path from the noise gets challenging as the power of the NLoS path increases which, in turn, results in higher range estimation error. 
\figref{fig:DelayVsAmplitude_h0} illustrates an intriguing phenomenon: 
the high-power NLoS path in the TX-RX channel significantly impacts the range estimation compared to a similar NLoS path in the Tag-RX channel. This occurs because the carrier signal over the TX-RX channel is used as a reference signal for ranging. Specifically, the high-power NLoS path in the TX-RX channel introduces errors in the OFDM symbol timing, which also affects the TX-Tag-RX channel, thereby compounding the error.
In practice, this scenario is unlikely to occur because, in any geometrically consistent channel, a high-power NLoS cluster would also introduce an apparent delay in the backscatter channel. In such cases, the range estimation error would result from the difference in the lengths of the dominant NLoS paths, which is not ideal for the requirements of \ac{LoS} path-based baseline ranging methods.
\subsection{Communication and Ranging Performance vs. SNR}
In this section, we present the limits on the communication and ranging performance of the proposed bi-static backscatter system as a function of the SNR. To capture the effects of signal impairments such as time and frequency offsets, we ensure that all units in the emulation setup are unsynchronized. 
\begin{figure*}[t]
\centering
\subfloat[BER vs. SNR\label{fig:noise_impact_BER}]{\includegraphics[width=.85\columnwidth]{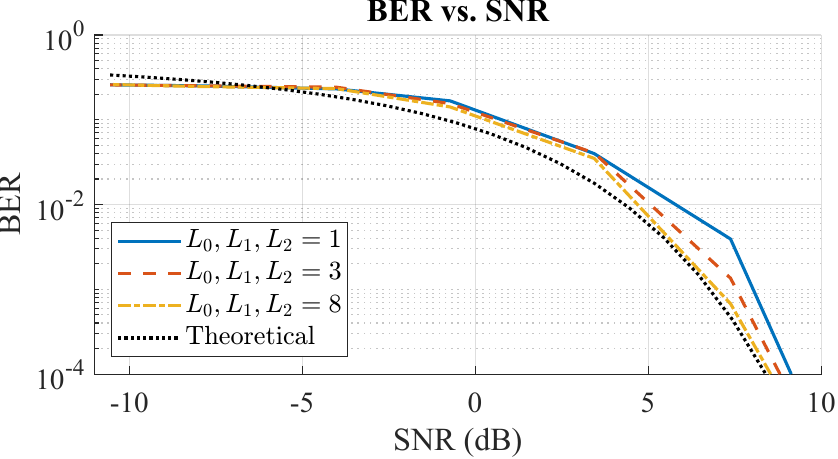}}\quad\quad
\subfloat[\ac{RMSE} of Ranging error (m) vs. SNR\label{fig:noise_impact_range} ]{\includegraphics[width=.85\columnwidth]{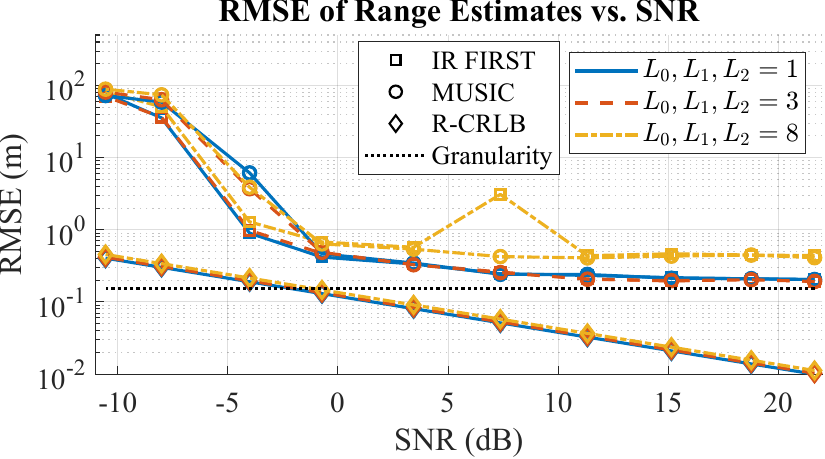}}
\caption{Impact of SNR on the communication and ranging performance of the bi-static backscatter system: (a) The BER performance is same as the achievable performance with any active radio with BPSK modulation. (b) \ac{RMSE} of the range estimates using IR First and MUSIC saturates to the granularity level $c/(N'\Delta F)$. The performance gap between IR First and MUSIC, and R-CRLB suggests the need of sophisticated methods for ranging in a multipath environment.\vspace{-0.1in}}
\label{fig:noise_impact}
\end{figure*}

\begin{figure}[t]
\centering
\includegraphics[width=.85\linewidth]{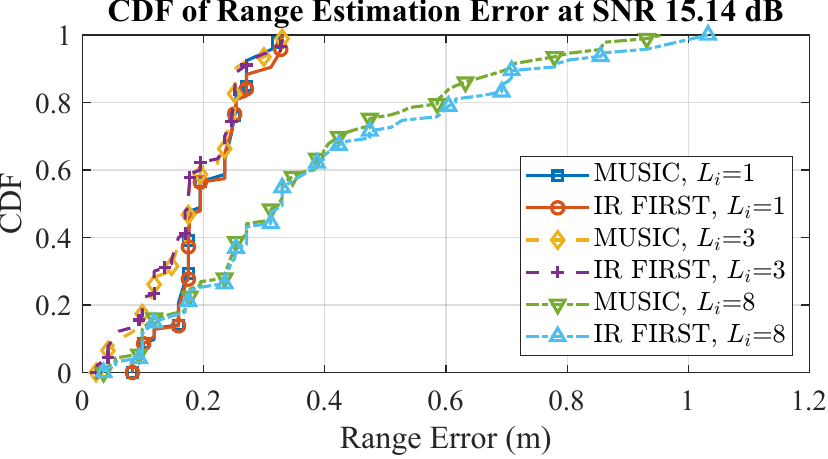}
\caption{Experimental Backscatter Ranging CDF: The number of multipath components in the environment severely affects the performance of IR First and MUSIC.}
\label{fig:ranging_cdf}
\end{figure}
To adjust the SNR of the backscatter signal, we control the attenuation level of the Tag-RX channel ($h_2$). At the receiver, it is essential to map the attenuation level to the baseband SNR. Our SNR estimation process at each attenuation level involves two stages. First, we deactivate the tag and capture the received baseband signal in the backscatter channel. We then demodulate the OFDM symbol to estimate the total power at the indices of utilized subcarriers, which represents the noise power $\hat{P}_N$ at the specified attenuation level. Subsequently, we activate the tag in CW mode and apply the same methodology to determine the total power in the active subcarriers, which now includes both signal and noise $\hat{P}_{S,N} = \hat{P}_S + \hat{P}_N$. Finally, we estimate the SNR by calculating the ratio of the signal power to the noise power, scaled by the ratio of the tag symbol duration to the OFDM symbol duration 
\begin{align}
    \hat{\mathrm{SNR}} = \frac{\hat{P}_S}{\hat{P}_N}~\frac{T_\text{sym}}{T_\text{OFDM}} = \left(\frac{\hat{P}_{S,N}}{\hat{P}_N} - 1\right) ~\frac{T_\text{sym}}{T_\text{OFDM}}.
\end{align}
The above equation leads to an important observation: the tag SNR per bit increases as the ratio between the tag symbol and OFDM symbol duration increases. In other words, accumulating multiple OFDM symbols results in processing gain for the tag demodulation. This is also intuitive, since for single-carrier BPSK, $E_b/N_0$ increases with the bit duration.

To measure the communication performance of the system, we use \ac{BER} of the tag packet as a metric. For this, we put the tag in the burst mode. The receiver follows the method described in \secref{sec:reader} to process the illuminator and backscatter signals, estimate the channels, detect the tag packet, estimate the channel observed by the tag symbols, and demodulate the tag packet symbols. We then calculate BER for each tag packet by comparing the demodulated tag packet symbols with the ground truth tag packet. 

To measure the ranging performance of the system, we use \ac{RMSE} of the range estimate as a metric. We put the tag in a CW mode. While the tag in the burst mode would provide similar ranging performance, we use the CW mode to avoid the inaccuracies in packet detection, a feature of communication, to affect the ranging performance and allow an independent analysis of both communication and sensing features. The receiver receives both illuminator signal and the backscatter signal and uses the IR First method with $N'=4096$ as described in \secref{sec:ranging} and the \ac{MUSIC} method to produce a range estimate for each measurement. Note that we use 4096 points to produce the pseudo-spectrum using \ac{MUSIC} algorithm, which yields the same range granularity as IR First.

For each attenuation level (and the associated SNR level) and the number of channel paths, we generate 20 scenarios with varying tag and scatterers' positions, resulting in 20 instances of TX-RX, TX-Tag and Tag-RX channels. For each scenario, we vary the attenuation level from 12 dB to 48 dB with 4 dB steps and collect 5 measurements for each attenuation level. Finally, we average the BER and determine the \ac{RMSE} of range estimates across all measurements for a given attenuation level and the number of channel paths.

\figref{fig:noise_impact_BER} shows the BER performance along with the expected theoretically achievable BER. As the figure shows, Tag-RX communication link is virtually similar to a generic low-rate BPSK communication link with an active radio. Therefore, it shows that there is no penalty in terms of communication performance for using backscatter tags with BPSK modulation compared to an active tags with similar modulation. 

\figref{fig:noise_impact_range} shows the \ac{RMSE} of range estimate along with the stochastic root-CRLB (R-CRLB) as a function of SNR. As the figure shows, the IR First and MUSIC methods in an environment with small number of multipath components achieve the granularity limit defined by $c/(N'\Delta F)$. Furthermore, as the number of paths in the channels, i.e., $L_0, L_1, L_2$ increases, the \ac{RMSE} of both estimation methods increases while the R-CRLB remains almost unchanged, which suggests the possibility of better ranging methods for heavy multipath (rich scattering) environments. 
This gap is further reinforced by \figref{fig:ranging_cdf} which shows the \ac{CDF} of the absolute errors in the range estimates for different numbers of multipath components in the channels. As the number of multipath components increases in the channel, the range estimation error also increases for both algorithms. 


%% file: crlb.tex
\subsection{CRLB of TX-RX range estimation}\label{app:Tx-Rx}

We start with the log-likelihood function 
\begin{align}
\ln \mathcal{L}(d_0, \mathbf{d}^0_\mathrm{NLoS}) 
&= - N \ln \pi - \ln \det(\mathbf{Q}_0) \nonumber \\
&\quad - \big(\hat{\boldsymbol{\mathcal H}}_0 - \boldsymbol{\mu}_0\big)^H
\mathbf{Q}_0^{-1}
\big(\hat{\boldsymbol{\mathcal H}}_0 - \boldsymbol{\mu}_0\big).
\end{align}
and its first-order derivative w.r.t.  $d_0$
\begin{align}\label{eq:diff-log-likelihood_0}
\frac{\partial}{\partial d_0} \ln \mathcal{L}(d_0, \mathbf{d}^0_\mathrm{NLoS})
= \frac{2B}{c} \, \re{
\big( \hat{\boldsymbol{\mathcal{H}}}_{1,2}^{+} - \boldsymbol{\mu}_0 \big)^{H}
\mathbf{Q}_0^{-1}
\frac{\partial \boldsymbol{\mu}_0}{\partial \tau_0}
}.
\end{align}
Differentiating \eqref{eq::diff-log-likelihood} once more w.r.t. $d_0$ yields 
\begin{align}
\frac{\partial^2}{\partial d_0^2} \ln \mathcal{L}(d_0, \mathbf{d}^0_\mathrm{NLoS})&=  \frac{2B}{c} \, \re{
\big( \hat{\boldsymbol{\mathcal{H}}}_{1,2}^{+} - \boldsymbol{\mu}_0\big)^{\H}
\mathbf{Q}_0^{-1}
\frac{\partial^2 \boldsymbol{\mu}_0}{\partial \tau_0^2 }
} \nonumber \\
&-\frac{2B}{c}\,
\frac{\boldsymbol{\mu}_0}{\tau_0}^{\H}
\mathbf{Q}_0^{-1}
\frac{\partial \boldsymbol{\mu}_0}{\partial \tau_0 }.
\end{align}
Using the relation $\mathbb{E}\left\{\hat{\boldsymbol{\mathcal{H}}}_{1,2}^{+} - \boldsymbol{\mu}_0\right\}=\boldsymbol{0}$, the Fisher information becomes
\begin{align}
\mathcal{I}(d_0)&= \frac{2B}{c} \, 
\frac{\boldsymbol{\mu}_0}{\tau_0}^{\H}
\mathbf{Q}_0^{-1}
\frac{\partial \boldsymbol{\mu}_0}{\partial \tau_0 }.
\end{align}

Considering \eqref{eq:mu-0} and assuming $\mathbf{Q}_0=\sigma_0^{2}\mathbf{I}_N$, we rewrite the term inside the expectation in \eqref{eq:FI_0_again} as
\begin{align}\label{eq:fisher0}
\frac{\partial \boldsymbol{\mu}_{0}}{\partial \tau_0}^{\H}
\mathbf{Q}_0^{-1} \:
\frac{\partial \boldsymbol{\mu}_0}{\partial \tau_0 } 
= \frac{1}{\sigma_0^2} \left\| \frac{\partial \boldsymbol{\mu}^+}{\partial \tau_0^0}\right\|_2^2,
\end{align}
where
\begin{equation} \label{eq:partial_derivative_again}
\frac{\partial \boldsymbol{\mu}_0}{\partial \tau_0^{0}}
=  \frac{1}{2} \exp[{\jmath (\phi_\text{RX}-\phi_\text{TX})}]
\boldsymbol{\mathcal{D}}_\zeta^0 
\frac{\partial \boldsymbol{\boldsymbol{\mathcal{H}}}_0}{\partial \tau_0^0},
\end{equation}
and $\boldsymbol{\mathcal{D}}_\zeta^0$ is given in \eqref{eq:D_zeta_0}.
Channel matrix $\boldsymbol{\boldsymbol{\mathcal{H}}}_0$ is given in \eqref{eq:multipath_ch01_discrete_freq_resp_tau_l} and its partial derivative w.r.t. $\tau_0^{0}$ is
\begin{equation} \label{eq:partial_derivative_once_more}
\frac{\partial \boldsymbol{\boldsymbol{\mathcal{H}}}_0}{\partial \tau_0^0} 
= -\jmath 2 \pi \alpha_{0}^{0}\exp \Big(-\jmath 2 \pi \frac{F_{\mathrm{c}}}{B} \tau_0^{0}\Big) \mathbf{f}(\tau_0^{0}) \odot\Big(\frac{F_{\mathrm{c}}}{B} \mathbf{1}_N+\frac{1}{N} {\mathbf n}\Big) 
\end{equation}
where 
$\mathbf{f}(\tau)=[\exp(\jmath \pi\frac{N-1}{2}\tau),\ldots,\text{exp}(-\jmath \pi\frac{N-1}{2}\tau) ]^{\T}$, and 
${\mathbf n}=[-(N-1)/2,\ldots,(N-1)/2]^\T$.
Inserting the expression of partial derivatives 
\eqref{eq:partial_derivative_once_more} and \eqref{eq:partial_derivative_again} 
into \eqref{eq:fisher0}, and simplifying the expression yields the final expression  \eqref{eq:CRLB_final_0}.
\subsection{CRLB of TX-Tag-RX range estimation}\label{app:Tx-Tag-Rx}
We  have 
\begin{align}
\ln \mathcal{L}(d_{1,2},\mathbf{d}_\text{NLoS}^{1,2}) 
&= - N \ln \pi - \ln \det(\mathbf{Q}_{1,2}) \nonumber \\
&\quad - \big(\hat{\boldsymbol{\mathcal H}}_{1,2}^{+} - \boldsymbol{\mu}^{+}\big)^H
\mathbf{Q}_{1,2}^{-1}
\big(\hat{\boldsymbol{\mathcal H}}_{1,2}^{+} - \boldsymbol{\mu}^{+}\big).
\end{align}
and 
\begin{align}\label{eq::diff-log-likelihood}
\frac{\partial}{\partial d_{1,2}} 
\ln \mathcal{L}(d_{1,2},\mathbf{d}_\text{NLoS}^{1,2})
= \frac{2B}{c} \, \re{
\big( \hat{\boldsymbol{\mathcal{H}}}_{1,2}^{+} - \boldsymbol{\mu}^{+} \big)^{H}
\mathbf{Q}_{1,2}^{-1}
\frac{\partial \boldsymbol{\mu}^{+}}{\partial \tau_{1,2}}
}.
\end{align}
Differentiating \eqref{eq::diff-log-likelihood} with respect to  $d_{1,2}$ yields 
\begin{align}
\frac{\partial^2}{\partial d_{1,2}^2} \ln \mathcal{L}(d_{1,2},\mathbf{d}_\text{NLoS}^{1,2})&=  \frac{2B}{c} \, \re{
\big( \hat{\boldsymbol{\mathcal{H}}}_{1,2}^{+} - \boldsymbol{\mu}^{+}\big)^{\H}
\mathbf{Q}_{1,2}^{-1}
\frac{\partial^2 \boldsymbol{\mu}^{+}}{\partial \tau_{1,2}^2 }
} \nonumber \\
&-\frac{2B}{c}\,
\frac{\boldsymbol{\mu}^{+}}{\tau_{1,2}}^{\H}
\mathbf{Q}_{1,2}^{-1}
\frac{\partial \boldsymbol{\mu}^{+}}{\partial \tau_{1,2} }.
\end{align}
Using the relation $\mathbb{E}\left\{\hat{\boldsymbol{\mathcal{H}}}_{1,2}^{+} - \boldsymbol{\mu}^{+}\right\}=\boldsymbol{0}$, the Fisher information  boils down to
\begin{align}
\mathcal{I}(d_{1,2})&= \frac{2B}{c} \, 
\frac{\boldsymbol{\mu}^{+}}{\tau_{1,2}}^{\H}
\mathbf{Q}_{1,2}^{-1}
\frac{\partial \boldsymbol{\mu}^{+}}{\partial \tau_{1,2} }.
\end{align}
Considering \eqref{eq:mu-plus} and assuming  $\tau_1^0=\beta \tau_{1,2}$ and $\tau_2^0=(1- \beta )\tau_{1,2}$, from the chain rule, we obtain
\begin{equation} \label{eq:chain-rule1}
\frac{\partial \boldsymbol{\mu}^{+}}{\partial \tau_{1,2}} 
= \beta \frac{\partial \boldsymbol{\mu}^{+}}{\partial \tau_1^0 } +(1-\beta)\frac{\partial \boldsymbol{\mu}^{+}}{\partial  \tau_2^0}.
\end{equation}
We further assume  $\mathbf{Q}_{1,2}=\sigma_{1,2}^{2}\mathbf{I}_N$, which helps us to rewrite the term in \eqref{eq:FI_12_again} as
\begin{align}\label{eq:fisher1}
\frac{\partial \boldsymbol{\mu}^{+}}{\partial \tau_{1,2}}^{\H}
\mathbf{Q}_{1,2}^{-1} \:
\frac{\partial \boldsymbol{\mu}^{+}}{\partial \tau_{1,2} } 
= \frac{1}{\sigma_{1,2}^2} \left\| \beta \frac{\partial \boldsymbol{\mu}^+}{\partial \tau_1^0} + (1-\beta)\frac{\partial \boldsymbol{\mu}^+}{\partial \tau_2^0} \right\|_2^2
,
\end{align}
where $\beta=\tau_1^0/\tau_{1,2}$.
The partial derivatives of $\boldsymbol{\mu}^{+}$ w.r.t. $\tau_1^{0}$ and $\tau_2^{0}$ respectively, are
\begin{align} \label{eq:chain-rule2}
\frac{\partial \boldsymbol{\mu}^{+}}{\partial \tau_1^{0}}
&=  + \frac{\jmath}{\pi} \exp[{\jmath (\phi_\text{RX}^\mp-\phi_\text{TX})}]
\boldsymbol{\mathcal{D}}_\zeta^0 \operatorname{diag}(\boldsymbol{\mathcal{G}}^{+})
\left [
\frac{\partial \boldsymbol{\boldsymbol{\mathcal{H}}}_1}{\partial \tau_1^0} \odot \boldsymbol{\mathcal{H}}_2^{+} \right ], \\
\frac{\partial \boldsymbol{\mu}^{+}}{\partial \tau_2^{0}}
&=  + \frac{\jmath}{\pi} \exp[{\jmath (\phi_\text{RX}^\mp-\phi_\text{TX})}]
\boldsymbol{\mathcal{D}}_\zeta^0 \operatorname{diag}(\boldsymbol{\mathcal{G}}^{+}) \left[
\boldsymbol{\mathcal{H}}_1 \odot \frac{\partial \boldsymbol{\mathcal{H}}_2^{+}}{\partial \tau_2^0} \right].
\end{align}
where $\boldsymbol{\mathcal{D}}_\zeta^0$ is given in \eqref{eq:D_zeta_0}.
Channel matrices $\boldsymbol{\boldsymbol{\mathcal{H}}}_1$ and $\boldsymbol{\boldsymbol{\mathcal{H}}}_2^+$  are given in \eqref{eq:multipath_ch01_discrete_freq_resp_tau_l} and \eqref{eq:multipath_ch2_discrete_freq_resp_tau_l}, respectively, and their partial derivatives w.r.t. $\tau_1^{0}$ and $\tau_2^{0}$ respectively, are
\begin{align}
\frac{\partial \boldsymbol{\boldsymbol{\mathcal{H}}}_1}{\partial \tau_1^0} 
=& -\jmath 2 \pi \alpha_{1}^{0}\exp \Big(-\jmath 2 \pi \frac{F_{\mathrm{c}}}{B} \tau_1^{0}\Big) \mathbf{f}(\tau_1^{0}) \odot\Big(\frac{F_{\mathrm{c}}}{B} \mathbf{1}_N+\frac{1}{N} {\mathbf n}\Big) \label{eq:partial_derivative_tau1} \\
\frac{\partial \boldsymbol{\mathcal{H}}_2^{+}}{\partial \tau_2^0}
=& -\jmath 2 \pi \alpha_2^{0}\exp\Big(-\jmath 2\pi \frac{F_c+F_{\text{shift}}}{B}\tau_{2}^0\Big)\mathbf{f}(\tau_2^{0}) \nonumber\\
& \odot\Big( \frac{F_{\mathrm{c}}-F_{\text{shift}}}{B} \mathbf{1}_N+\frac{1}{N} {\mathbf n}\Big).
\label{eq:partial_derivative_tau2}
\end{align}
The channel matrices $\boldsymbol{\boldsymbol{\mathcal{H}}}_1$ and $\boldsymbol{\boldsymbol{\mathcal{H}}}_2^+$ in \eqref{eq:multipath_ch01_discrete_freq_resp_tau_l} and \eqref{eq:multipath_ch2_discrete_freq_resp_tau_l} and their partial derivatives in \eqref{eq:partial_derivative_tau1}--\eqref{eq:partial_derivative_tau2} may then be inserted into \eqref{eq:CRLB_final_12} to get the final CRLB expression for the backscatter range estimate.